\documentclass[a4paper]{article}

\usepackage[a4paper,left=2cm,right=2cm,top=2cm,bottom=2.5cm]{geometry}

\usepackage{amsmath}
\usepackage{graphicx}
\usepackage{hyperref}

\usepackage{verbatim}
\usepackage[bbgreekl]{mathbbol}
\usepackage{color}
\definecolor{light-gray}{gray}{0.975}
\usepackage{xcolor}

\usepackage{listings}
\newcommand*\Suppressnumber{
    \let\thelstnumber\relax
}
\newcommand*\Reactivatenumber[1]{
  \setcounter{lstnumber}{\numexpr#1-1\relax}
    \let\thelstnumber\origthelstnumber
    \refstepcounter{lstnumber}
}

\newcommand{\xb}{\ensuremath{\textbf{x}}}
\newcommand{\vb}{\ensuremath{\textbf{v}}}
\newcommand{\eb}{\ensuremath{\textbf{e}}}
\newcommand{\bb}{\ensuremath{\textbf{b}}}
\newcommand{\jb}{\ensuremath{\textbf{j}}}
\newcommand{\X}{\ensuremath{\textbf{X}}}
\newcommand{\V}{\ensuremath{\textbf{V}}}
\newcommand{\Eb}{\ensuremath{\textbf{E}}}
\newcommand{\Bb}{\ensuremath{\textbf{B}}}
\newcommand{\du}{\ensuremath{\text{d}}}

\newcommand{\MM}{\ensuremath{\mathbb{M}}}
\newcommand{\C}{\ensuremath{\mathbb{C}}}
\newcommand{\LaB}{\ensuremath{\mathbb{\Lambda}}}

\definecolor{red}{rgb}{0.89411764705882357, 0.10196078431372549, 0.10980392156862745}
\definecolor{blue}{rgb}{0.21568627450980393, 0.49411764705882355, 0.72156862745098038}
\definecolor{green}{rgb}{0.30196078431372547, 0.68627450980392157, 0.29019607843137257}
\definecolor{purple}{rgb}{0.59607843137254901, 0.30588235294117649, 0.63921568627450975}
\definecolor{orange}{rgb}{1.0,                 0.49803921568627452, 0.0                }
\definecolor{yellow}{rgb}{1.0,                 1.0,                 0.2                }
\definecolor{brown}{rgb}{0.65098039215686276, 0.33725490196078434, 0.15686274509803921}
\definecolor{pink}{rgb}{0.96862745098039216, 0.50588235294117645, 0.74901960784313726}
\definecolor{grey}{rgb}{0.6, 0.6, 0.6}
\begin{document}

\title{Evaluation of performance portability frameworks for the
implementation of a particle-in-cell code}

\author{
Victor Artigues$^{1,2}$,
Katharina Kormann$^1$,
Markus Rampp$^2$,
Klaus Reuter$^{2,*}$
\\
\\
$^{1}$Max-Planck-Institut f{\"u}r Plasmaphysik,\\
Boltzmannstr.~2, 85748 Garching, Germany\\
\\
$^{2}$Max Planck Computing and Data Facility\\
Gießenbachstr.~2, 85748 Garching, Germany\\
\\
$^*$Correspondence: \texttt{klaus.reuter@mpcdf.mpg.de}
% ---
}

% --- original stylefile
%\institute{
%Max Planck Computing and Data Facility\\
%Gießenbachstraße 2, Garching, 85748, Germany\\
%\email{
%Email: \{luka.stanisic, klaus.reuter\}@mpcdf.mpg.de
%}
%}
% ---

% To run latexdiff you need to comment the following maketitle!
\maketitle

\begin{abstract}
This paper reports on an in-depth evaluation of the
performance portability frameworks Kokkos and RAJA with respect to their suitability for the
implementation of complex particle-in-cell (PIC) simulation codes, extending
previous studies based on codes from other domains.
At the example of a particle-in-cell model, we
implemented the hotspot of the code in C++ and parallelized it using OpenMP, OpenACC,
CUDA, Kokkos, and RAJA, targeting multi-core (CPU) and graphics (GPU) processors.
Both, Kokkos and RAJA appear mature, are usable for complex codes,
and keep their promise to provide performance portability across different
architectures.
Comparing the obtainable performance on state-of-the art hardware, but also
considering aspects such as code complexity, feature availability, and
overall productivity, we finally draw the conclusion that the Kokkos
framework would be suited best to tackle the massively parallel
implementation of the full PIC model.
\end{abstract}

\section{Introduction}

Modern high-performance computing (HPC) systems are getting
increasingly complex, in particular concerning the hardware
architecture of the clusters' nodes.
This trend is reflected in the development of the ``Top 500'' ranking
of the fastest supercomputers \cite{top500}, where a growing fraction of
HPC systems deploys various types of accelerators or
coprocessors, in addition to the prevailing multi-core
CPUs. As of November 2018, such
accelerated systems contribute a total (peak-performance
weighted) share of about 40\% within the Top 500 \cite{top500} list, with more than
two dozens of different types of accelerators (GPUs in the majority of
systems). For an individual HPC system, in particular at the high
end of the list, GPUs contribute a large fraction of the nominal
peak performance.

While for handling inter-node communication the Message Passing
Interface (MPI \cite{MPI}) remains unchallenged as \emph{the}
programming model which has proven stable, reliable, portable, and well
supported over more than 25 years, a de-facto standard
for programming heterogeneous compute nodes has yet to emerge.
The situation is quite adequately described by the term ``MPI + X'' which has
been around for a number of years \cite{MPIpX}, but today, the ``X''
still represents a variety of node-level programming paradigms which
are mostly specific for a certain type of hardware, or even vendor.

With OpenMP\cite{OpenMP}, OpenACC\cite{OpenACC}, and OpenCL\cite{OpenCL}, to name the most relevant and widespread ones, there
is a set of language extensions to C and Fortran available that---
at least partly---offer portable programming across various types of
compute nodes:
OpenMP\cite{OpenMP} has been very successful for programming
multi-core CPUs, and most recent specifications target also
accelerators. However, there is
currently only very limited compiler support for accelerators,
and some of the semantics differ between CPUs and accelerators.

Conceptually similar to OpenMP, OpenACC\cite{OpenACC} was designed
as a high-level, directive-based approach to GPU programming and
has been quite successful, in particular as an alternative to CUDA,
but also supporting multi-core processors. In practice, however, there is
only limited compiler support for OpenACC (proprietary compilers by PGI and
CRAY, and experimental support in GCC). For complex application codes,
both, the OpenMP, and---maybe to
a slightly lesser degree---the OpenACC-based approach likely require
target-specific adaptations of data structures and flow control in
order to achieve good performance on both, CPUs and GPUs.

OpenCL\cite{Munshi2011, OpenCL} was designed for heterogeneous systems with
CPUs and GPUs, but, in practice, has limited support by compilers and
runtimes. Moreover, it exposes many of the complexities of the GPU
hardware architecture to its programming model. In particular,
parallelism is expressed in the form of so-called kernels which are
mapped to the CPU's or GPU's threads in a SIMT or SIMD fashion.
This often poses severe challenges to HPC codes which typically
express parallelism in loops or---more fashionably---in complex task
graphs, both of which requires a major rewriting of code, and, for
Fortran programs, the construction of C interfaces.

In the absence of a de-facto standard for achieving portable performance,
scientific HPC application development, especially when targeting
high-end machines, is facing an enormous challenge. In practice
multiple code versions, code paths, or alike, need to be implemented
for the same algorithm, employing different node-level programming
models specialized for the target hardware platforms. Applying advanced software engineering
techniques, some of the complexity can be encapsulated in
certain ``lower'' layers of the software architecture of an
application (e.g.~in a custom domain-specific language
\cite{Fuhrer:2014,Clement:2018}), but readability, maintainability,
and sustainability of such code is often significantly impacted.

Hence, there is significant demand and motivation to develop tools for
abstracting the source code from the hardware. Performance portability
frameworks enable programmers to target multiple architectures, ideally
achieving good performance on any of them without the need to implement and
optimize individually.
Existing frameworks typically build upon the C++ programming language
with templates (ideally, architecture-dependent decisions are taken 
automatically at compile time) and provide a limited set of building
blocks for parallelism while
hiding the complexity of the target architecture from the application programmer. The
promise is that a \emph{single} source code can run efficiently on multiple architectures.
Following the paradigm of ``separation of
concerns'' the framework eventually delegates compilation and
execution to well-established ``native'' programming
models and runtimes, such as the aforementioned OpenMP (for multi-core
CPUs), or the proprietary CUDA model (for NVIDIA GPUs).
Those ``backends'' can be implemented and maintained by
specialists in an application-agnostic way and thus become transparent for the
application programmer.
For a more comprehensive review on
options for performance portability we refer to the introduction of
Ref.~\cite{Albany2019} and the references therein.

In this paper, we shall focus on Kokkos \cite{Kokkos} and RAJA
\cite{RAJA} which, at the time of writing, are the two
leading C++ performance portability frameworks. Both use advanced
meta-programming techniques to generate architecture-specific code and optimizations
at compile time and are freely available as open source under
BSD-type licenses.
Similar, though still in beta state and therefore not considered currently, is
the alpaka performance portability library\cite{ZenkerAsHES2016,
MathesP3MA2017}. It provides hardware abstraction to support single-source
accelerator development.
Specifically, this paper reports on our experiences with employing
Kokkos and RAJA for achieving performance portability of a
complex particle-in-cell (PIC) code across various types of
multi-core and GPU-accelerated HPC platforms, and compares it to
platform-optimized versions of the code. In addition to the
computational performance, our assessment considers aspects such as code
complexity, feature availability, and overall productivity of
the approach.

The paper is structured as follows. 
Subsection \ref{sec:related} gives a brief review of related work on
the assessment of performance-portability frameworks.
Next, subsection \ref{sec:method} introduces the methodology and the goals of the
present work and subsection \ref{sec:gempic} briefly introduces the
numerical model of our study.
In section \ref{sec:perf_port_framework}, we briefly summarize the
main concepts of Kokkos
and RAJA and put them into context by means of code examples from our
application.
We then turn towards a usability review of both frameworks in section
\ref{sec:user_review}, before we report in detail on the performance
achieved on state-of-the art CPU and GPU nodes in section
\ref{sec:benchmark_results}.
Finally, the paper closes with a summary and conclusions in section
\ref{sec:conclusion}.

\subsection{Related work}
\label{sec:related}

To our knowledge, there exist only few comparisons of Kokkos, RAJA,
and other parallel frameworks at the level of a complex HPC application.

Martineau et al.~did a comparison of Kokkos, RAJA, OpenACC, OpenMP 4.0, CUDA,
and OpenCL based on the Tealeaf application, a miniapp solving the heat
conduction equation using finite differences\cite{Martineau, Martineau_bis}.
The authors report a 5\% to 30\% performance penalty for Kokkos and RAJA
compared to architecture-specific implementations.

Sunderland et al.~report on the Kokkos refactoring of the large legacy code
base Uintah, used to model turbulent combustion \cite{UINTAH2016}.
In their case, the introduction of Kokkos views and proper memory layouts even
increased the performance of specific kernels thanks to more efficient memory
accesses and vectorization.

More recently, Demeshko et al.~report specifically on a Kokkos port of
parts of the complex finite element framework Albany in great
detail\cite{Albany2019}. They focus on the refactoring process, and present
performance comparisons on the target platforms CPU, GPU, and KNL, though a
baseline defined by the original code's performance before the porting is
lacking.

\subsection{Methodology}
\label{sec:method}

We consider a complex HPC application from the plasma physics domain,
based on a numerical model which solves the
Vlasov--Maxwell system of equations using a Particle-In-Cell (PIC)
method\cite{GEMPIC}. The fact that a full-scale implementation of the
code has yet to be developed and optimized for state-of-the-art HPC systems, which is a
multi person-year effort, served as the main motivation for
the pilot-study and assessment of portability frameworks presented
here.

As a baseline implementation, we extracted a generic part from the original
FORTRAN implementation taken from SeLaLib \cite{selalib} and rewrote it in
C++. Starting out from this code, six different parallel versions were
developed, namely, using plain OpenMP directives, plain OpenACC directives,
plain CUDA, Kokkos, hybrid OpenMP/Kokkos, and RAJA. The baseline version and
the six parallel versions, targeting multi-core and graphics processors, were
extensively benchmarked, and are compared with respect to their performance. In
addition, we address \emph{soft factors} such as the programmers'
productivity, the code complexity, and the overall experience.

Since the fundamental computational kernels of the PIC method are somewhat
complementary to the patterns encountered in finite-element methods or
mesh-based domain-decomposition schemes 
that were targeted in the
aforementioned studies, our assessment may serve as another major
building block for judging the relevance of portability frameworks
for HPC application development as a whole.

\subsection{Numerical model}
\label{sec:gempic}

The particle-in-cell method solves a hyperbolic conservation law by representing a distribution function by macro-particles that evolve in a Lagrangian frame along the characteristic equations accociated with the conservation law. These macro-particles are represented by their position in phase space and a weight. Often the particle description is coupled to a field description of some moments of the distribution function. We study the solution of the Vlasov--Maxwell equations in six-dimensional phase space. The Vlasov equation models the evolution of a species $s$ of a plasma in its self-consistent and external electromagnetic fields
\begin{equation}\label{eq:vlasov}
\partial_t f_s(\xb, \vb, t) + \vb \cdot \nabla_{\xb} f_s(\xb, \vb, t) + \frac{q_s}{m_s} \left( \Eb(\xb,t) + \vb \times \Bb(\xb,t) \right)\cdot \nabla_{\vb} f_s(\xb, \vb, t) = 0,
\end{equation}
where $f$ denotes the distribution function, $\Eb$ the electric, and $\Bb$ the magnetic field, and $q_s$ and $m_s$ the charge and mass of the particles, respectively.
The self-consistent fields can be computed from Maxwell's equations
\begin{subequations}\label{eq:vlasov_maxwell_equations}
\begin{align}
&\frac{\partial \Eb}{\partial t} - \nabla \times  \Bb = - \jb, \quad
\label{eq:maxwell_ampere} \\
&\frac{\partial \Bb}{\partial t}  + \nabla  \times \Eb = 0, \label{eq:maxwell_faraday}
\\
&\nabla \Eb = \rho, \quad \label{eq:gauss}\\
&\nabla \Bb = 0, \label{eq:grad_b}
\end{align}
\end{subequations}
where $\rho$ and $\jb$ denote the charge and current density, respectively, which are defined as velocity moments of the distribution functions,
\begin{equation}
\rho = \sum_sq_s \int f_s \du \vb, \text{   and   } \jb = \sum_s q_s \int \vb f_s \du \vb.
\end{equation}
Particle methods represent the distribution function $f_s$ by a large
number of macroparticles that are defined by their position $(\xb_p, \vb_p)$
in six-dimensional phase space and a (constant) weight, and evolve over time. The fields are represented on a grid by some discretization method.

In our case study, we follow the geometric electromagnetic particle-in-cell (GEMPIC) scheme as proposed in \cite{GEMPIC} where the fields are represented by conforming spline finite elements as proposed in \cite{Buffa:2011}. The common parallel structure of the particle-in-cell method are particle loops which are embarrassingly parallel but assemble and evaluate grid-based quantities that are shared between several processes and thus require a reduction step (and the solution of the field equations) following the particle loop. Our implementation focuses of a subset of equations, namely a particle loop that identifies the position of the particle in the grid, evaluates the magnetic field at the particle position, updates $x_{1,p}$ and $v_{2,p}$, $v_{3,p}$, and accumulates the first component of the current density. The particle loop is then followed by a reduction of the current density. Appendix \ref{sec:gempic_detailed} gives a more detailed description of the implemented method. Moreover, we refer to \cite{GEMPIC} where this subset of instructions is presented as the operator $H_{p_1}$. 

Although the study uses a particular operator stemming from the GEMPIC discretization, we believe that the findings are general enough to apply to other types of particle-in-cell schemes.
Note that we focus in the present work only on shared-memory parallelization
with a simple data structure storing particle position, velocity, and weight in an
array-of-structures data type. Including advanced
data types that enable vectorization and efficient data access
(cf.~e.g.~Ref.~\cite{OptimalData}) and adapting them
to the special requirement of the current-deposition loop as it appears in the
GEMPIC equations is a different topic which we do not address in this work.

\section{Performance Portability Frameworks}
\label{sec:perf_port_framework}

The aim of performance portability frameworks like Kokkos and RAJA is to enable scientific
programmers to write generic parallel code that can be compiled and run
on several parallel architectures while minimizing or even eliminating the
need to implement architecture-specific code. At the same time, nearly
the same computational performance as obtained from an
architecture-specific implementation is supposed to be achieved.
Thus, the application programmer has to take care of only a single
code version and is shielded from technical details of different
target architectures. Moreover, the code is expected to run at high performance
also on future HPC hardware, provided that the chosen performance-portability
framework is properly maintained.

To point out the concepts, similarities, and differences of Kokkos and RAJA, we
present a recap of both frameworks, partly based on code from our
implementations of the numerical model.
For direct comparison of the source code from the different programming models
we have included a color-coded listing in appendix \ref{appendix:colorcoded}.
These code excerpts transport the essential ideas behind Kokkos and RAJA. We
used Kokkos version 2.7.00 and RAJA version 0.6.0 during development work.
For further details on the abstractions and the programming models, we
refer the reader to the official documentation of Kokkos\cite{KokkosGuide}
and RAJA\cite{RAJAGuide}.

In the following sections, we will briefly address the role of the individual
building blocks for the implementation of our model code. The key feature of
both Kokkos and RAJA is to offer abstractions for the memory (organized in
so-called views) and the parallel operations. Kokkos defines the following six
fundamental abstractions:
\begin{enumerate}
\setlength\itemsep{-0.33em}
\item \emph{execution spaces} specify on which processor to execute,
\item \emph{execution patterns} specify the parallel operation (i.e.~for, reduction, scan, task),
\item \emph{execution policies} specify how to execute the pattern,
\item \emph{memory spaces} specify where to allocate memory and store data,
\item \emph{memory layouts} control the mapping of indices to physical memory,
\item \emph{memory traits} specify how to access the memory.
\end{enumerate}
RAJA supports various types of parallel operations, such as loops (for), reductions,
atomics (handled by the data structure in Kokkos), and scans, and defines a \emph{policy} for each of them specifying both
where and how the parallel operation is executed. The memory abstraction is
through \emph{views} that wrap the pointer to the actual memory and enable
processing of the data independent of its index organization that is  defined by
the concept of \emph{layouts}. Note that Kokkos automatically allocates the
memory according to the \emph{memory space} while RAJA requires the user to
explicitly define the memory layout for each case.
These abstractions enable the implementation of parallel code as outlined
in the following.

%\subsection{Multidimensional array}
\subsection{Array types and memory management}

Kokkos and RAJA both offer multidimensional array primitives called
\emph{views}. These views allow for an abstraction of the storage layout from
the data which, most importantly, relieves the programmer from
architecture-specific optimization work.
For example, on cache-based architectures such as CPUs, it is of advantage to
access memory linearly (i.e.~multiple consecutive elements from each thread)
whereas on GPUs it typically performs better to access memory in a coalesced fashion
(i.e.~one element only per lightweight thread).
Using the view abstraction, the optimal layout and padding can be
chosen automatically and individually for each architecture (\emph{execution
space}).

In both frameworks, a view contains only metadata which is stored in host
memory, plus a pointer to the actual data that can be located on the host or
on the device.
Kokkos allocates memory together with views, whereas RAJA does not allocate
memory.
To be able to, e.g., define an array independent of its location, Kokkos
implements the so called \emph{HostMirror} type.
This special view transparently guarantees data access to device memory
from the host. In our code, the names of such views contain the keyword
\texttt{\_host}.
With RAJA, managing host and device memory allocation and transfers is the
responsibility of the user, similarly to the CUDA heterogeneous programming
model.

In the codes presented in the appendix, Kokkos views are used, e.g., in the
lines 101-104 of the listing \ref{appendix:colorcoded},
\begin{lstlisting}[numbers=none]
|\textcolor{red}{this->view\_j\_dofs\_local = Kokkos::View<double*>("view\_j\_dofs\_local", this->part\_mesh\_coupling.view\_n\_dofs\_host(0));}|

|\textcolor{red}{this->view\_j\_dofs\_local\_host = Kokkos::create\_mirror\_view(this->view\_j\_dofs\_local);}|

\end{lstlisting}
\vspace{-1em}
\begin{lstlisting}[firstnumber=101]
|\textcolor{red}{for(long i=0; i<this->view\_j\_dofs\_local\_host.size(); i++) \{}|
	|\textcolor{red}{this->view\_j\_dofs\_local\_host(i) = 0.0;}|
|\textcolor{red}{\}}|
|\textcolor{red}{Kokkos::deep\_copy(this->view\_j\_dofs\_local, this->view\_j\_dofs\_local\_host);}|
\end{lstlisting}
and RAJA views are used, e.g., in the lines 108-122 of the listing \ref{appendix:colorcoded}:
\begin{lstlisting}[numbers=none]
|\textcolor{green}{this->d\_j\_dofs\_local = memoryManager::allocate<double>(this->part\_mesh\_coupling.n\_dofs);}|

|\textcolor{green}{this->raja\_j\_dofs\_local.~RajaVector1DType();}|
|\textcolor{green}{new(\&this->raja\_j\_dofs\_local) RajaVector1DType(this->d\_j\_dofs\_local, this->part\_mesh\_coupling.n\_dofs);}|
\end{lstlisting}
\vspace{-1em}
\begin{lstlisting}[firstnumber=108]
|\textcolor{green}{\#if defined(RAJA\_ENABLE\_CUDA) \&\& defined(I\_USE\_CUDA)}|
	|\textcolor{green}{RAJA::forall<RAJA::cuda\_exec<32{>}>(RAJA::RangeSegment(0, this->part\_mesh\_coupling.n\_dofs),}|
	 |\textcolor{green}{[*this] RAJA\_DEVICE (int i) \{}|
			|\textcolor{green}{if(i<this->part\_mesh\_coupling.raja\_n\_dofs(0)) \{}|
				|\textcolor{green}{this->raja\_j\_dofs\_local(i) = 0.0;}|
			|\textcolor{green}{\}}|
	 |\textcolor{green}{\});}|
|\textcolor{green}{\#else}|
	|\textcolor{green}{RAJA::forall<RAJA::omp\_parallel\_for\_exec>(RAJA::RangeSegment(0, this->part\_mesh\_coupling.n\_dofs),}|
	 |\textcolor{green}{[this] (int i) \{}|
			|\textcolor{green}{if(i<part\_mesh\_coupling.raja\_n\_dofs(0)) \{}|
				|\textcolor{green}{raja\_j\_dofs\_local(i) = 0.0;}|
			|\textcolor{green}{\}}|
	 |\textcolor{green}{\});}|
|\textcolor{green}{\#endif}|
\end{lstlisting}

\subsection{Parallel patterns}
\label{section:Execution patterns}

Kokkos and RAJA both implement data parallel operations (for, reduce, scan),
and Kokkos in addition also a task parallel operation (task).
These parallel operations are called \emph{patterns} in the case of Kokkos,
whereas RAJA subsumes them under the term \emph{policy}, see section
\ref{section:Execution policies} below, however the concepts are the same.
The calls to parallel code can be done via lambda functions in both frameworks.
With Kokkos, a functor can be used alternatively to wrap parallel code.

The inputs of the lambda function or the functor have to match the expected input parameters
of the execution pattern and policy. This prohibits the use of arguments to pass data,
therefore lambda functions seem easier to work with on a small scale.
To get access to the data, functors require all the data as class members.

According to our experience, functors are often easier to test and somewhat more readable,
especially for source codes with many lines.
However, plain functors seem to pose a problem when multiple parallel functions
to be accessed via the same functor need to be implemented.
To address this, Kokkos uses an \emph{Execution Tag}.
The tag is passed as a template parameter to the \emph{execution policy},
which will then feed it to the \textit{operator()},
making the specialization to call a certain internal function a compile-time decision.
An execution tag is used at line 176 of Listing \ref{appendix:colorcoded}:
\begin{lstlisting}[firstnumber=176]
|\textcolor{red}{auto policy = Kokkos::TeamPolicy<pic\_routine, ExecSpace>}|
\end{lstlisting}
The first template parameter, \texttt{pic\_routine}, is the tag that will define which
\textit{operator()} to use. The second parameter is the \emph{execution
space} defining, at compile time, the information on where to execute the
code.

At the time of writing and for both the frameworks, the reduction loop
supported on both CPU and GPU only covered scalar reductions. Vector
reduction is a crucial feature for the implementation of a PIC method
such as the one considered by us in the present paper.
While RAJA does
not provide a ``ready-to-use'' solution for vector reduction, Kokkos provides
it by the use of a \emph{scatter\_view}. The \emph{scatter\_view} deals with
the allocation of per-thread memory for thread-safe access to the vector,
see the example below taken from the lines 316, 445 of the listing \ref{appendix:colorcoded}.
For completeness we have added a call to the contribute function which performs the reduction:
\begin{lstlisting}[firstnumber=445]
|\textcolor{red}{ViewScatterAccessType scatter\_access = scatter\_view.access();}||\Suppressnumber||\Reactivatenumber{366}|
|\textcolor{red}{scatter\_access(index1d) += view\_j1d(i) * splinejk;}|
\end{lstlisting}
\vspace{-1em}
\begin{lstlisting}[numbers=none]
|\textcolor{red}{Kokkos::Experimental::contribute(hamiltonian\_splitting.view\_j\_dofs\_local, scatter\_view);}|
\end{lstlisting}
For RAJA, we have implemented our own vector reduction by using a different view for each particle's
partial result, and summing the partial results at the end. See the example below adapted from
lines 200, 447 and 214 of the listing \ref{appendix:colorcoded}.
\begin{lstlisting}[firstnumber=200]
|\textcolor{green}{//RAJA CPU: initialisation of the handmade reduction 2D array}|
|\textcolor{green}{RAJA::kernel<fdPolicy>(RAJA::make\_tuple(RAJA::RangeSegment(0, total\_num\_threads)),}|
	|\textcolor{green}{[\&,this] (int i\_part) \{}|
		|\textcolor{green}{for(int i=0; i<part\_mesh\_coupling.n\_dofs; i++)}|
			|\textcolor{green}{raja\_private\_raja\_array(i\_part).raja\_handmade\_reduce(i) = 0.0;}|
	|\textcolor{green}{\});}|
|\Suppressnumber||\Reactivatenumber{446}|
|\textcolor{green}{raja\_private\_raja\_array(i\_part).raja\_handmade\_reduce(index1d) += util\_raja.raja\_j1d(i) * splinejk;}|
|\Suppressnumber||\Reactivatenumber{213}|
|\textcolor{green}{//RAJA: CPU handmade reduction}|
|\textcolor{green}{RAJA::kernel<fdPolicy>(RAJA::make\_tuple(RAJA::RangeSegment(0, part\_mesh\_coupling.n\_dofs)),}|
	|\textcolor{green}{[\&,this] (int i\_part) \{}|
	|\textcolor{green}{for(int i=0; i<total\_num\_threads; i++)}|
		|\textcolor{green}{d\_j\_dofs\_local[i\_part] += raja\_private\_raja\_array(i).raja\_handmade\_reduce(i\_part);}|
	|\textcolor{green}{\});}|
\end{lstlisting}

\subsection{Execution policies}
\label{section:Execution policies}

To specify how a parallel pattern is executed, Kokkos knows two types
of \emph{execution policies}, the \emph{RangePolicy} and the \emph{TeamPolicy}.
The range policy simply defines a range of indices on which a function will
be called. No assumption can be made about the order of the concurrent execution,
and it is not allowed to synchronize the threads.
The team policy adds additional features on top of the range policy. It enables
hierarchical parallelism and scratch memory.
The scratch memory is a very important feature for the present PIC
application, as each thread needs private variables to compute updated
particle positions, velocities, etc.
At line 176 of Listing \ref{appendix:colorcoded}, a TeamPolicy, with shared\_size bytes
per-thread of scratch memory is declared.
\begin{lstlisting}[firstnumber=176]
|\textcolor{red}{Kokkos::TeamPolicy\{...\}.set\_scratch\_size(1,Kokkos::PerThread (shared\_size));}|
\end{lstlisting}

RAJA decides on which hardware the code will run based on the template
arguments of the policy.
The indices for the data-parallel operation are defined using RangeSegments,
a RAJA class to define a range $[\![n, m-1]\!]$. Moreover, e.g., ranges with
constant strides $\{n, n+m, n+2*m, ...\}$ are possible. In line 189 of listing
\ref{appendix:colorcoded}, we launch a lambda function on the GPU with 256
threads per thread-block on the indices $[\![0, n\_particles-1]\!]$, as shown
below.
\begin{lstlisting}[firstnumber=189]
|\textcolor{green}{RAJA::forall<RAJA::cuda\_exec<256{>}>(RAJA::RangeSegment(0,}|
	|\textcolor{green}{this->particles.group[0].n\_particles),}|
		|\textcolor{green}{[=,*this] RAJA\_DEVICE (int i\_part)\{}|
			|\textcolor{green}{\{operator\_RAJA\_GPU\}}|
		|\textcolor{green}{\});}|
\end{lstlisting}
These segments can further be combined into lists of segments for more
complex access patterns. However, this advanced index management feature is
not necessary for our PIC model and has not been tested in this study.

In the following we turn towards a review of the different frameworks, based
on our experience from implementing the PIC algorithm.

\section{Usability review}
\label{sec:user_review}

\subsection{Implementation overview and methodology}
\label{section:implementations}

In order to compare Kokkos and RAJA with OpenACC and the architecture-specific
programming models,
we developed multiple versions of the PIC operator.
\begin{description}
\item[C++] A standalone C++ reference implementation of the PIC
routine was written first. This code includes input/output and driver functions to run, time, and verify
the computation, and serves as the starting point for
the other codes below.
\item[OpenMP] Starting from the C++ code, we implemented a thread-parallel
version for the CPU using plain OpenMP. Results from this code define the baseline
to compare to the Kokkos-CPU and RAJA-CPU codes, see below.
\item[Kokkos] Next, the parallel loops from the OpenMP code were refactored
using Kokkos and were verified to produce correct results on CPUs and GPUs. In
the following, we refer to this version as Kokkos-CPU when run on the CPU
with the Kokkos-OpenMP back end, and as Kokkos-GPU when run on the GPU with the
Kokkos-CUDA back end, respectively.
\item[OpenMP/Kokkos] Related to the pure Kokkos implementation, we developed
a hybrid code version which uses OpenMP directives in combination with Kokkos
views, with the motivation to identify the overhead of the pure Kokkos
implementation. Moreover, such hybrid codes naturally emerge during code
refactoring work.
\item[RAJA] A parallel implementation was developed using RAJA, and
verified on CPUs and GPUs. In the following, we refer to this version as
RAJA-CPU when run on the CPU with the RAJA-OpenMP back end, and as RAJA-GPU
when run on the GPU with the RAJA-CUDA back end, respectively.
\item[CUDA] To define a baseline for Kokkos' and RAJA's GPU
performance, a plain CUDA version was implemented.
\item[OpenACC] To complement this comparison, a plain OpenACC version
was implemented, which runs on the CPU and on the GPU.
\end{description}
To compile all but the last code we used GCC 6.3 and CUDA 9.1 with appropriate library
and compiler flags to optimize for the specific hardware. In particular, the
flags \texttt{-O3 -march=native -fopenmp} were passed to gcc, and, e.g., the flags \texttt{-O3
-arch=sm\_70} were passed to nvcc with the \texttt{-arch} flag
matching the actual GPU architecture, NVIDIA Volta in this example. For Kokkos, e.g., the macro
\texttt{KOKKOS\_ARCH="SKX,Volta70"} was specified in addition to specify the target
platforms, here Intel Skylake CPU and NVIDIA Volta GPU.
To compile the OpenACC code we used PGI 19 with the \texttt{-fast -acc}
optimization flags and the respective target specialization flags for the CPU
(\texttt{-ta=multicore}) and GPU (e.g., \texttt{-ta=tesla:cc70}).
% Update: We don't use managed memory any more.
%Note that for the latter, we intentionally rely on CUDA Unified Memory in
%order to include OpenACC as a high-level performance portability approach,
%handling memory transfers and, in particular, deep copies of nested data
%structures conveniently behind the scenes.

In the following, we discuss the implementation experience with Kokkos and
RAJA individually and address soft factors such as code complexity and
productivity, before we turn towards performance benchmarks in section
\ref{sec:benchmark_results}.

\subsection{Kokkos implementation}
\label{section:review_kokkos}

To perform a refactoring of existing sequential C++ code into parallel Kokkos
code, essentially a two-step process is necessary in the most simple case:
First, replace legacy C-style array allocations or, similarly, higher-level
data structures such as \emph{std::vector} with \emph{Kokkos::View} types,
and, second, replace \emph{for} loops with \emph{parallel\_for} and move the
loop bodies into functors or lambda functions.

As mentioned previously, Kokkos execution patterns offer two possibilities to
run user code in parallel.
For the present work, the functor approach was preferred to the lambda
functions, allowing for easy testing and leading to well-readable code.

As a feature of convenience and robustness, Kokkos provides a default
execution space that will consider architectures in the order CUDA, OpenMP,
pthreads, and serial, if available. This default execution space makes it
optional to the programmer to specify the architecture onto which the code is
going to be deployed. Complex implementations can of course have parallel
sections specified to use distinct execution spaces.

Vector reductions are a crucial feature for our application. The Kokkos
reference only covers scalar reductions in the
documentation\footnote{https://github.com/kokkos/kokkos/wiki/Custom-Reductions:-Build-In-Reducers,
accessed on 09/11/2018}, examples, and test files.
However, Kokkos provides a ScatterView class under the experimental namespace
that can be used to implement vector reductions.
It has a slightly different behavior on the CPU and the GPU:
On the CPU, the ScatterView class will duplicate its View internally per
thread. After the parallel section, the \textit{contribute} function has to
be called explicitly to compute the reduction from all the internal Views.
On the GPU, the ScatterView does not duplicate its View, hence, value updates
from concurrent threads need to be performed atomically.
On present GPU hardware, the atomic add operation is efficiently
implemented\cite{LectureAtomic}, which makes it a good choice to perform a
reduction instead of using variable duplication.
The ScatterView class expects a template parameter to request to use the
duplication or the atomic access, but there is no default setting. Hence, we
specified different template settings for CPU and GPU using a preprocessor
macro. In case of the Kokkos framework, this was the only time two versions of
a code section (though 2 lines only) had to be implemented.

Lastly, we need some scratch memory for each thread. For the Kokkos
implementation, this is simply achieved by declaring all the needed variables
inside the parallel section, such that allocation is done per thread.
The total size of the scratch memory inside a parallel loop needs to be known
before the parallel dispatch is launched, similarly to CUDA shared memory.
Different memory levels with different size caps, corresponding to L1 and L2
cache sizes, can be accessed on the CPU, and similarly for shared memory on
the GPU.

\subsection{RAJA implementation}
\label{section:RAJA}

Turning towards RAJA and the refactoring of existing sequential C++ code, the
same two-step process as already discussed with Kokkos is necessary: First,
replace array types with \emph{RAJA::View} types, and, second replace
\emph{for} loops with \emph{forall} and move the loop bodies into lambda
functions. It is the duty of the user to allocate memory for the RAJA views.

A difference to note is that RAJA does not implement defaults, e.g., for
architectures, but rather forces the programmer to consider and specify all
the necessary settings explicitly. On the source code level this requires
branches to specialize for CPU and GPU versions.

Second, RAJA does not support functors for the parallel dispatch but only lambda
functions which required to structure the code differently, compared to Kokkos.

As already discussed for the Kokkos implementation, scratch memory is needed,
for which RAJA offers an implementation of shared memory on the GPU.
On the CPU, we use regular views.

RAJA does not offer ready-to-use vector reductions, however, it was straight
forward to implement such reductions based on views.

\subsection{Qualitative comparison}

\begin{table}[h]
\begin{center}
\caption{Qualitative comparison of the programming models based on a subjective
ranking on the scale low, medium, high, as experienced by the programmers.}
\label{tab:comparison}
\begin{tabular}{|| c | c c c c c ||}
\hline
criterion     & OpenMP  & OpenACC & CUDA  & Kokkos  & RAJA   \\
\hline\hline
code clarity  & high    & high    & low   & medium  & medium \\
productivity  & high    & medium  & low   & medium  & medium \\
portability   & low     & medium  & low   & high    & high   \\
performance   & high    & high    & high  & high    & medium \\
\hline
\end{tabular}
\end{center}
\end{table}

In the following, a qualitative review on the code complexity, programmer's
productivity, portability, and performance is given, summarized in table
\ref{tab:comparison}.

The directive-based OpenMP and OpenACC programming models were the least intrusive when
applied to the loops of the PIC routine, starting from the sequential C++
code. Kokkos and RAJA required both significant restructuring of the existing
code for the parallel dispatch via functors or lambda functions. CUDA
required a comparable amount of rewriting effort, in particular to map the loops
onto a CUDA grid of threads and thread blocks. The overhead for OpenMP and OpenACC in
terms of lines of code is the smallest, followed by Kokkos. For RAJA, the
overhead is about a factor two compared to Kokkos in many places since
separate code for CPU and GPU can be necessary. The CUDA version is
comparable to the Kokkos code in terms of lines of code.

Concerning the programmer's productivity, it took a graduate-level C++
programmer about two months to learn and apply the Kokkos framework successfully
to implement the PIC routine. RAJA, conceptually similar and tackled with
having the knowledge of Kokkos, took about another month. The OpenMP and CUDA
implementations were done faster due to previous experience.
To develop the OpenACC code, it was necessary to do the implementation in
analogy to the CUDA implementation, keeping the correspondence of OpenACC
gangs and CUDA thread blocks in mind.
Remarkably, Kokkos
and RAJA keep their promise of basically providing a GPU version for free based
on implementation work mainly done using the CPU. For completeness, we include
the performance ranking in table \ref{tab:comparison} but refer to the following
section \ref{sec:benchmark_results} for details.
Moreover, it should be noted that the specification of the target platform in
the Kokkos Makefile via the \texttt{KOKKOS\_ARCH} variable already enables the
use of correct compiler optimization flags for the respective platform,
while for the other frameworks the user has to set these flags manually,
a procedure which is tedious and prone to errors.

A drawback common to both frameworks, Kokkos and RAJA, is the fact that the
hardware-specific code generated at compile time via C++ template meta
programming is not accessible to the programmer for inspection. This limitation
does not only affect Kokkos and RAJA, but is well-known to any C++ programmer
who uses templates. There is no way to obtain the resulting code in the
high-level C++ or C++/CUDA language, rather the programmer has to work at the
level of assembly code.

\subsection{Hybrid OpenMP/Kokkos}

Even though performance portability frameworks enable portability of the
code, they often come with a certain overhead which we have also observed in
our application (cf. the following section). For this reason, we were
interested in the question whether this overhead is intimately connected to
the data structures or rather to the implementation. For this reason, we have
modified the Kokkos code such that it uses OpenMP \emph{parallel for} directives
instead of the Kokkos loops but keeps the Kokkos \textit{views} as array
objects.

Because Kokkos does reference counting, accessing a regular view from a pure
OpenMP parallel section would cause significant overhead. In particular, any
operation on a view would be an atomic operation. If one wants to use a view
from conventional OpenMP parallel code, the trait of the view needs to be set
to Unmanaged. This is what we used for the hybrid OpenMP/Kokkos code, see
e.g.~the lines 13 and 136 of listing \ref{appendix:colorcoded}:
\begin{lstlisting}[firstnumber=13]
|\textcolor{orange}{ViewVector2DUnmanagedType view\_particle\_array\_unmanaged;}||\Suppressnumber||\Reactivatenumber{135}|
|\textcolor{orange}{this->view\_i\_weight\_unmanaged = this->view\_i\_weight;}|
\end{lstlisting}
With these modifications it was possible to eliminate the overhead for
Kokkos as will be reported on in the next section. We did not perform such
an experiment for RAJA. However, since memory management is more transparent in
RAJA, no difficulties should arise in this case.

\section{Performance benchmarks}
\label{sec:benchmark_results}

\subsection{Hardware}
\label{sec:hardware}

For the performance benchmarks presented in the following we consider a
single multi-core CPU socket and a single GPU. This is commonly considered a
fair comparison when measuring the performance of heterogeneous codes.
We used the following hardware in the scope of this work.
\begin{description}%[leftmargin=8em,style=nextline]
\item[IvyBridge-Kepler] Single node with two Intel Xeon E5-2680 v2 @2.80GHz CPUs
(IvyBridge) and two NVIDIA K40m GPUs (Kepler, PCIe 3.0) \cite{K40m}.
\item[POWER8-Pascal] Single node with two IBM POWER8 processors and four NVIDIA
Tesla P100 GPUs (Pascal, NVLINK) \cite{P100}.
\item[Skylake-Volta] Single node with two Intel Xeon Gold 6148 2.4GHz CPUs
(Skylake) and two NVIDIA V100 GPUs (Volta, PCIe 3.0) \cite{V100}.
\end{description}
The IvyBridge-Kepler system was used for the majority of the development
work. To obtain performance numbers on state-of-the art hardware, we turned
towards running the code on the POWER8-Pascal and the Skylake-Volta systems.

\subsection{Preparatory work}

After porting, we performed analysis and optimization work on all codes for a comparable
amount of time, hence, the numbers presented below should provide a
representative relative picture of the performance one may expect from the
various approaches, when starting from a sequential C++ PIC code. No specific optimizations were necessary for the various versions of the code, except for a false sharing issue encountered with the RAJA version.
We first consider a single process and run 3 iterations to measure the single
core performance of the PIC routine. We choose the number of particles to be
10 million on a 16 by 8 by 8 mesh (i.e.~$N_g=16 \cdot 8 \cdot 8$) and use splines of degree 3 (and 2) which we
keep constant for all the runs discussed in the following. With these
parameters, the plain C++ code runs on a single IvyBridge core in about $7.5$
seconds per iteration.

The PIC routine implements the particle-to-mesh transfer for which the
spline basis for the finite element description of the fields needs to be
evaluated at the particle position---or integrated over the particle
trajectory in the time step. This requires the localization of the particles
within the mesh, which is done via modulo operations.
A straight-forward optimization was to cache the modulo computations for each
iteration, thereby replacing computation with memory storage and reuse. This
reduced the processing time by about a factor of 2 for all the
implementations.

Note that we did not change the fundamental data structure used to store the
particles. We use a C++ class (struct), hence, the ensemble of particles is
stored as an array of structures. Alternatively, one could consider using a
structure of arrays or a mix of both, potentially improving the performance
due to better vectorization. However, finding optimal particle data
structures for PIC codes is still the subject of ongoing research, cf. e.g.,
Ref.~\cite{OptimalData}, and is beyond the scope of the present work.

\subsubsection{False sharing mitigation}
\label{section:false_sharing}

During the development work it was found that the RAJA-based code shows the
worst scaling on the CPU, which is due to sub-optimal memory access, leading
to false sharing of cached data between threads. The LIKWID performance tool
\cite{LIKWID} was used successfully to shed light on the cause.
Running a ``FALSE\_SHARING'' analysis revealed that the OpenMP-based code has
about 200 MB of last level cache (LLC) hits when using 10 cores whereas the
Kokkos code only has about 10 MB. However, for the RAJA-CPU implementation, a
much larger value of 18 GB was determined.
This metric refers to the amount of memory the processor has to synchronize
between cores via the last level cache (L3 cache), because data present in
the L1 or L2 caches of one core were modified by other cores at the same
time, a situation known as false sharing.

Ideally, threads on different CPU cores work on data that are far away in
memory compared to the size of a cache line, which is typically 8 doubles on
x86\_64 platforms.
Our algorithm and implementation retrieves the position, velocity and charge of
a particle from a large array. It then works with local variables to compute the
new state of the particle before updating the large array. With OpenMP, the
local variables are defined as thread private. With Kokkos-CPU, we use the per
thread scratch memory. Therefore, the memory allocated to each thread is
allocated as one block.

A very early version of our RAJA-CPU code used a simple $\text{\#threads} \times
3$ View for the local variables, leading to false sharing, see listing \ref{code:false1}.
The approach taken
to mitigate this situation is known as cache alignment or padding. In order
to avoid threads copying neighbor values into their caches, ghost values are
added between the values. Instead of having a $\text{\#threads} \times 3$
View, we allocate a $\text{\#threads} \times 8$ View where the first 3 values
are used as before and the next 5 are never touched. They serve as padding to
make sure the core only copies the relevant elements into its cache, and not
its neighbor's vector.
See the listing \ref{code:false2} for the modified code with padding.
Padding improves the scaling significantly,
however, drawbacks are higher memory requirements and transfers.
The final and best solution is to create a situation similar to OpenMP and
Kokkos-CPU, which allocate local variables as contiguous blocks per thread.
To do so, we regrouped all the memory needed per thread in order to avoid any false sharing,
as shown in listing \ref{code:false3}.
This implementation scales moderately well, as reported in the next section.

\begin{lstlisting}[
caption={Code version with false sharing.},
label={code:false1}]
|\textcolor{green}{this->d\_xi = memoryManager::allocate<double>(N\_THREADS*3);}|
|\textcolor{green}{this->d\_vt = memoryManager::allocate<double>(N\_THREADS*3);}|

|\textcolor{green}{RAJA::forall<RAJA::omp\_parallel\_for\_exec>(RAJA::RangeSegment(0, N\_THREADS),}|
|\textcolor{green}{[*this] RAJA\_DEVICE (int i) \{}|
	|\textcolor{green}{this->raja\_private\_raja\_array(i).raja\_xi.~RajaVector1DType();}|
	|\textcolor{green}{new(\&this->raja\_private\_raja\_array(i).raja\_xi) RajaVector1DType(\&(this->d\_xi[i*3]), 3);}|
	|\textcolor{green}{this->raja\_private\_raja\_array(i).raja\_box.~RajaVector1DType();}|
	|\textcolor{green}{new(\&this->raja\_private\_raja\_array(i).raja\_box) RajaVector1DType(\&(this->d\_vt[i*2]), 2);}|
	|\textcolor{green}{\{...\}}|
|\textcolor{green}{\});}|
\end{lstlisting}

\begin{lstlisting}[
caption={Code version with padding.},
label={code:false2}]
|\textcolor{green}{this->d\_xi = memoryManager::allocate<double>(N\_THREADS*8);}|
|\textcolor{green}{this->d\_vt = memoryManager::allocate<double>(N\_THREADS*8);}|

|\textcolor{green}{RAJA::forall<RAJA::omp\_parallel\_for\_exec>(RAJA::RangeSegment(0, N\_THREADS),}|
|\textcolor{green}{[*this] RAJA\_DEVICE (int i) \{}|
	|\textcolor{green}{this->raja\_private\_raja\_array(i).raja\_xi.~RajaVector1DType();}|
	|\textcolor{green}{new(\&this->raja\_private\_raja\_array(i).raja\_xi) RajaVector1DType(\&(this->d\_xi[i*3]), 3);}|
	|\textcolor{green}{this->raja\_private\_raja\_array(i).raja\_box.~RajaVector1DType();}|
	|\textcolor{green}{new(\&this->raja\_private\_raja\_array(i).raja\_box) RajaVector1DType(\&(this->d\_vt[i*2]), 2);}|
	|\textcolor{green}{\{...\}}|
|\textcolor{green}{\});}|

\end{lstlisting}

\begin{lstlisting}[
caption={Code version with per-thread contiguous memory.},
label={code:false3}]
|\textcolor{green}{shared\_size = 1362;}|
|\textcolor{green}{this->memory\_pool = new double[N\_THREADS*shared\_size]();}|

|\textcolor{green}{RAJA::forall<RAJA::omp\_parallel\_for\_exec>(RAJA::RangeSegment(0, N\_THREADS),}|
|\textcolor{green}{[\&,this] (int i) \{}|
	|\textcolor{green}{int memory\_position = i*shared\_size;}|
	|\textcolor{green}{raja\_private\_raja\_array(i).raja\_xi.~RajaVector1DType();}|
	|\textcolor{green}{new(\&raja\_private\_raja\_array(i).raja\_xi) RajaVector1DType(\&(memory\_pool[memory\_position]), 3);}|
	|\textcolor{green}{memory\_position += 3;}|
	|\textcolor{green}{raja\_private\_raja\_array(i).raja\_vt.~RajaVector1DType();}|
	|\textcolor{green}{new(\&raja\_private\_raja\_array(i).raja\_vt) RajaVector1DType(\&(memory\_pool[memory\_position]), 2);}|
	|\textcolor{green}{memory\_position += 2;}|
	|\textcolor{green}{\{...\}}|
|\textcolor{green}{\});}|
\end{lstlisting}

\subsection{Performance results}
\label{section:performance}

\begin{figure*}[tb]
\centering
\includegraphics[width=\textwidth]{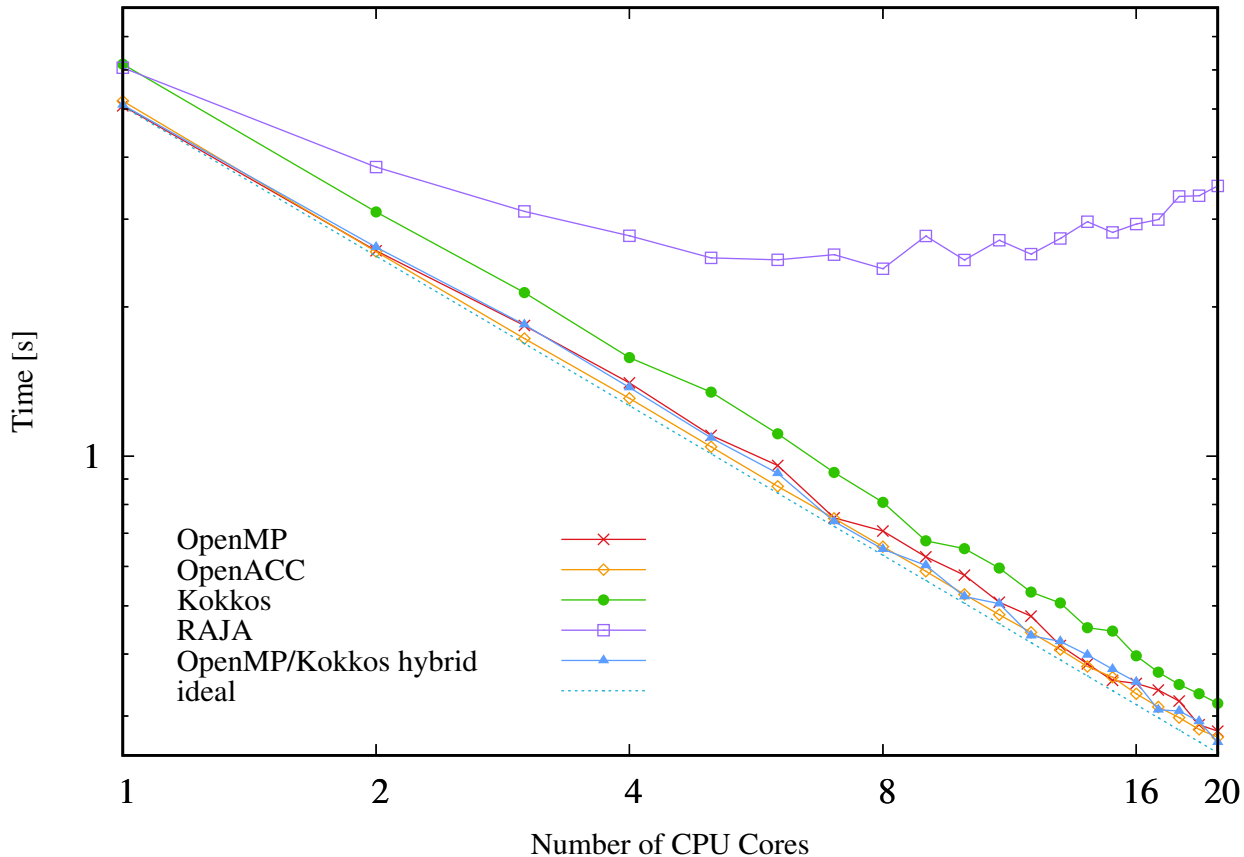}
\caption{Log-log plot of the compute times per iteration as functions of the number of CPU cores.
We compare the performance of Kokkos, RAJA, and OpenACC on the CPU to plain OpenMP. Moreover, the scaling curve of the hybrid
code that uses OpenMP directive-based loop-parallelism on data stored in
Kokkos views is shown.
The codes were run on a
20-core Intel Xeon Gold 6148 2.4GHz (Skylake)
CPU.
The times shown were averaged over 10 runs.}
\label{fig:cpu}
\end{figure*}

\begin{figure*}[tb]
\centering
\includegraphics[width=\textwidth]{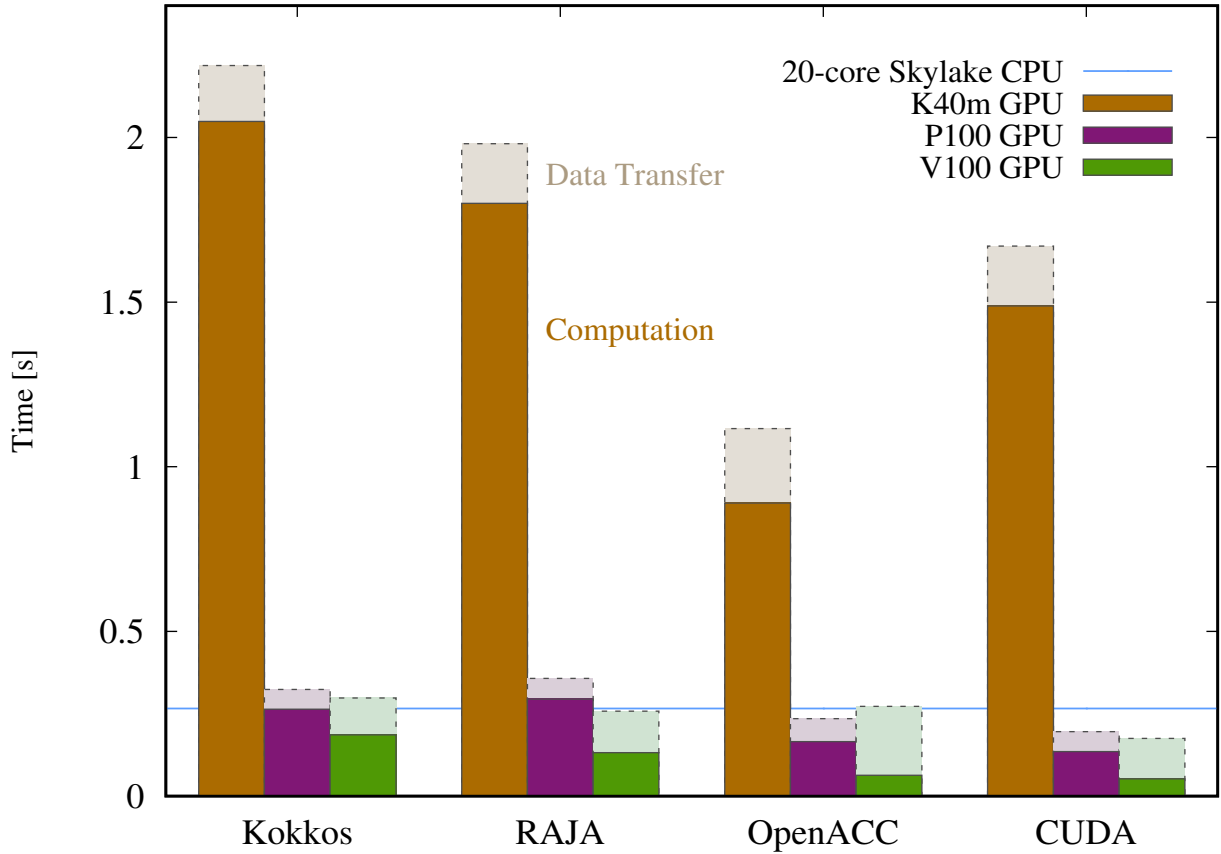}
\caption{Plot of the times per iteration for various GPU models.
We compare the performance of Kokkos, RAJA, and OpenACC to CUDA.  In addition to
the compute time the time necessary for one data transfer is shown.
The codes were run on a NVIDIA K40m, a NVIDIA P100, and a NVIDIA V100 GPU.
For direct comparison, a horizontal line indicates the best result obtained on the
20-core Intel Xeon Gold 6148 2.4GHz (Skylake) multicore
CPU, cf.~Fig.~\ref{fig:cpu}.
The times shown for the GPUs were averaged over 200 runs.}
\label{fig:gpu}
\end{figure*}

In this section, we provide a performance comparison of the different
implementations on the Skylake CPU, and the K40m, P100, and V100 GPUs,
representing state-of-the-art hardware at the time of writing
(cf.~Sec.~\ref{sec:hardware} for details on the hardware).
Fig.~\ref{fig:cpu} shows for each implementation in terms of the wall
clock time per iteration the parallel scaling on the CPU.
Fig.~\ref{fig:gpu} shows the
performance results from runs on the GPUs for comparison.

\subsubsection{CPU}
%
%Victor:
%OpenMP scaling (1 to 20): 18.15
%Kokkos scaling (1 to 20): 19.35
%RAJA scaling (1 to 20): 1.73
%RAJA scaling (best ratio at 10): 2.44
%Hybrid scaling (1 to 20): 19,18
%OpenACC scaling (1 to 20): 19,08

On the CPU, the OpenACC implementation turns out to be the fastest and scales
nearly ideally up to the full 20 Skylake cores with a speedup of about $19$.
Following closely, the next-ranked CPU code is the plain OpenMP implementation
which as well does show a near-ideal parallel scaling and speedup of about $18$.
%
%OpenMP is performing very similarly to OpenACC but is, average over the number
%of cores used, $4.6$\% slower and $2.8$\% slower on 20 cores.
Overall, OpenMP is performing very similarly to OpenACC, it is only about
$2.8$\% slower on 20 cores.
\footnote{Note that we used PGI to compile the OpenACC
code and GCC to compile the other codes, each with aggressive optimization flags
enabled (cf.~Sec.~\ref{section:implementations}).}
The Kokkos implementation, in comparison and averaged over the different core
numbers under consideration, is about a factor of $1.21$ slower than the OpenACC
code which we attribute to Kokkos-internal overhead.
Obviously, the performance and also the scaling of the RAJA code is the
worst in the present comparison, although the per-thread memory management
was already improved (cf.~Sec.~\ref{section:false_sharing}). Presumably,
this could be further optimized, however, this would require a
considerable performance tuning effort which is exactly what one wants to avoid when
choosing to build the implementation based on a performance portability framework.
The hybrid implementation which uses plain OpenMP directives in combination with
Kokkos views shows virtually identical performance as the plain OpenMP
implementation, with a parallel speedup of $19.18$ on 20 cores.  Hence, it is
possible to optimize certain critical parts of a code in situations when the
overhead from the Kokkos parallel execution is not acceptable.

\subsubsection{GPU}
Turning towards the GPUs, in addition to the compute time, we also have to
consider the time for data transfers between the host and the device.
%
%Such a transfer is necessary when the data needs to be synchronized between
%various processes.
In a full-fledged MPI-parallel PIC code, the operation on the particle data is embarrassingly
parallel and only needs synchronization
%between the processes
when particles
%need to be
are redistributed between processes due to particle sorting for reasons of data
locality. Other occasions for the exchange of particle data between host and
device are the initial setup or regular checkpointing. The frequency of such data
transfers depends on the characteristics of a particular setup.
%On the other hand,
Only the field data needs to be synchronized in each step, however,
%the data describing the fields
that data is almost negligibly small compared to the particle data.
For our study, we therefore measure separately the compute time and the time
needed for the data transfer, where we synchronized the data between host and device
in each time step. The total time in a realistic scenario will usually be close to
the pure compute time due to infrequent data transfers.

What concerns the data transfer time, it is important to first recall that the three GPU
models under consideration use different interconnects. The P100 is connected
via the NVLink communication bus with a transfer rate of 20 gigatransfers (GT)
per second, while the K40m and the V100 use a PCIe 3.0 bus with a transfer rate
of 8 GT/s. Therefore, the transfer times are roughly 2 times smaller with
NVLink, as can be clearly seen from the plot.
%Victor: this explains the P100 vs V100 difference for OpenACC!!!
%OpenACC, using Unified Memory, relies heavily on fast communications.
% Update: We don't use managed memory any more.
%In the OpenACC case with managed memory, we explicitly access the result
%arrays on the host CPU after the computation has finished in order to enforce a
%device-to-host memory transfer by the CUDA runtime. In the other cases,
%the transfer is done explicitly.
%
The compute and transfer times shown in Fig.~\ref{fig:gpu} were
determined using the NVIDIA \emph{nvprof} profiler.

%Victor:
%WITH TRANSFER
%K40m, compared to OpenACC
%Kokkos:+98.82%
%RAJA:+77.58%
%CUDA:+49.69%

%WITHOUT TRANSFER
%K40m, compared to OpenACC
%Kokkos:+130.26%
%RAJA:+102.30%
%CUDA:+67.32%
We now compare the
compute time %%% previously: total time
for the various implementations on the
three GPU models.
On the oldest GPU hardware, the K40m, the rank order is given by OpenACC,
CUDA, RAJA, and Kokkos. CUDA is about $67$\% slower than OpenACC. The RAJA
runs take a factor of $2.02$ and the Kokkos runs take a factor of $2.30$
longer than the times measured for OpenACC. This order changes on the more
recent hardware.
%
%Victor:
%WITH TRANSFER
%P100, compared to CUDA
%Kokkos:+65.54%
%RAJA:+82.81%
%OpenACC-GPU:+20.28%
%V100, compared to CUDA
%Kokkos:+69.96%
%RAJA:+46.99%
%OpenACC-GPU:+55.29%

%WITHOUT TRANSFER
%P100, compared to CUDA
%Kokkos:+95.38%
%RAJA:+118.44%
%OpenACC-GPU:+22.58%
%V100, compared to CUDA
%Kokkos:+253.09%
%RAJA:+150.65%
%OpenACC-GPU:+19.92%
On the P100 GPU, the CUDA code is the fastest, followed by OpenACC ($1.23$),
Kokkos ($1.95$), and RAJA, the latter being slower
than CUDA by a factor of $2.18$.

On the V100 GPU, again CUDA delivers the fastest result, followed by OpenACC
($1.20$), RAJA ($2.51$), and finally Kokkos which is a factor
of $3.53$ slower than CUDA. Note that in the OpenACC case, in particular,
the transfer time is larger on the V100 (PCIe) than on the P100 (NVLink),
which makes the OpenACC code overall run the fastest on the P100 GPU.
%
%
%CUDA stays the fastest, beating RAJA-GPU by a factor of $1.8$ on
%the P100 and $1.5$ on the V100. This ordering is reversed on the V100 where
%RAJA-GPU is the second slowest and Kokkos-GPU the third. Kokkos-GPU is slower
%than CUDA by about $65$ percent on the P100 and $70$ percent on the V100.
%Also on the V100, OpenACC-GPU is the slowest, by a factor of $2.6$ compared to CUDA.
%
Moreover, the compute time is by far the smallest on the V100 GPU
for the CUDA implementation, where the data transfer takes about a factor of
$2.3$ longer than the computation.
In comparison to the best result from the 20-core Skylake CPU, the computation
is significantly faster on the V100 in all cases.

The fact that the CUDA code becomes relatively faster when going to more recent
GPUs is caused by an implementation detail.  Atomic updates have received
significantly improved hardware support with recent GPUs.  Our CUDA
implementation uses atomic additions on a global array to perform the
reductions,
whereas our OpenACC implementation uses atomic updates on per-gang private array
copies followed by a final reduction across the gangs, reducing the concurrent
accesses compared to the CUDA approach and therefore showing relatively better
performance on the older K40m
GPU.  The
OpenACC per-gang solution was chosen because it demonstrated much better
performance on the multicore CPU compared to a solution with a single global
array, while showing rather similar performance on the GPUs.

%But the compute time gets smaller than
%the transfer time, by a factor of $2.3$ for CUDA on the V100, therefore
%giving smaller factors between the codes when comparing the sum of
%computational and transfer time.
%
%New questions: Why doesn't Kokkos improve more than 9%? Almost all the spee-up is compensated by the NVLink-PCIe transfer rate
%Note: CUDA is very fast compared to transfer time. Almost all the spee-up is compensated by the NVLink-PCIe transfer rate
%Victor:
%P100 to V100:
%RAJA:-31.31%
%Kokkos:-8.61%
%CUDA:-11.50%
%OpenACC:+20.50%
%This paragraph is now irrelevant
% \textcolor{red}{RAJA and CUDA show very little difference in compute time between the P100
% and the V100, by less than $2$ percent. The two codes are not using shared memory which
% could partly explain the lack of improvement: The V100 would be able to provide more
% shared memory at a higher bandwidth. Looking at the communication cost, the
% transfer of the data between the host is 2 times faster on the P100, we
% therefore obtain faster results on the P100 in most cases when considering
% the time to solution. Kokkos' GPU result is the only that improves when going
% from the P100 to the V100, by about 40\%, indicating that Kokkos does indeed
% perform hardware-specific optimizations at compile time.}

\section{Summary}
\label{sec:conclusion}

This paper presents a performance and usability assessment of two major
performance portability frameworks, Kokkos and RAJA, which are considered for
the future development of a high-performance C++ implementation of a
particle-in-cell approach to solve the Vlasov--Maxwell equations. We focus on
the node-local part of the computation, comparing generic implementations
based on Kokkos, RAJA, and OpenACC, to CPU- and GPU-specific implementations
based on OpenMP and CUDA, respectively.

\subsection{Usability assessment}

Considering programmability and usability, Kokkos and RAJA offer rather
similar concepts and levels of abstraction. Both frameworks provide generic
building blocks for parallel programming, targeting CPU and accelerator
platforms. Code specialization to a specific processor is done at compile
time based on C++ template meta programming, and is thus hidden from the user.
Regarding the features necessary for our PIC application such as vector
reductions and scratch memory, Kokkos offers all of them whereas for
RAJA some additional implementation work was needed.

Kokkos provides useful default values for the majority of relevant
template parameters. If not specified otherwise, the code is compiled for the
``fastest'' architecture available, as defined by the ordering CUDA, OpenMP,
pthreads, and serial.
With the default parameters, the execution space (CPU or GPU) of a parallel
section will automatically match the memory space (host memory or device
memory)
which significantly facilitates implementation and
simplifies the code. Moreover, with Kokkos
we managed avoiding architecture-specific preprocessor macros in the code. The Kokkos
project is very well documented and the developers are supportive on GitHub,
according to our experience.
As a plus, a simple architecture-specification string set by the user in the
Kokkos Makefile automatically enables the correct set of hardware-specific
optimization flags for the compiler.

RAJA does not implement such architectural default values for the template
parameters. Hence, the user has to specify the architecture-specific parameters
explicitly, e.g.~by employing preprocessor macros. RAJA required us to have
multiple of such branches in the code for CPU and GPU compilation. While this
adds some fine-grained control options, the underlying paradigm of writing a
single source code for multiple architectures gets somewhat compromised.

It should be noted that both the Kokkos-GPU and RAJA-GPU codes for
our PIC application were obtained ``for free'' in the sense that all
development started out on multi-core CPUs and no GPU-specific code
was ever written (except for some preprocessor branches to set template
parameters in the case of RAJA), thereby confirming the idea of portability
at good performance.

\subsection{Performance assessment}

In total, seven C++ versions of the PIC routine were developed, namely a
sequential one, and parallel codes using OpenMP, OpenACC, CUDA, Kokkos, hybrid
OpenMP/Kokkos, and RAJA.  The performance was measured on a Skylake multicore
CPU and on three NVIDIA GPUs from different hardware generations.  Only limited
effort was spent on optimizing each individual
implementation.

On the CPU, Kokkos shows acceptable overhead compared to plain OpenMP of about
14\% and scales nearly as well as plain OpenMP.  Moreover, we have also
demonstrated that the overhead of the Kokkos framework over a pure OpenMP
implementation is not linked to the use of Kokkos views for data management.  Therefore, it is possible
to mitigate the performance penalty of the Kokkos framework by manually
optimizing critical kernels.
The results we have obtained with RAJA
%by investing comparable effort
appear a bit inferior on the CPU since the parallel scalability is hampered
by a false sharing issue that could only partially be solved. On the other
hand, the performance on the GPU is comparable or even slightly better than
the one of Kokkos across different GPU hardware generations.
On state-of-the-art GPUs, CUDA turns out to be the fastest choice, followed by OpenACC, Kokkos and
RAJA.  Overall, OpenACC is very competitive on both, the CPU and the GPU, and should be kept
in mind as an option for performance portability, especially with improved
compiler support that is to be expected in the near future (GCC and LLVM, in
addition to PGI).

In summary, both Kokkos and RAJA seem mature, usable for production codes, and
keep their promise to provide performance portability across different hardware
architectures.

\bibliography{ppf.bib}{}

\appendix

\section{Numerical model}
\label{sec:gempic_detailed}

This section gives a more detailed description of the equations implemented in
our case study. The basis for particle-in-cell methods are the  characteristic equations associated to \eqref{eq:vlasov} are given by
\begin{equation}\label{eq:vlasov_char}
\frac{\du \X}{\du t} = \V, \quad \frac{\du \V}{\du t} = \left( \Eb(\X,t) + \V \times \Bb(\X,t) \right),
\end{equation}
along which the value of the distribution function remains constant over time. 
Particle methods represent the distribution function by a number $N_p$ of macro-particles that evolve according to the characteristic equations. The macro-particles are characterized by their position in phase space and a weight, that is particle $p$ is represented by $(\xb_p, \vb_p, w_p)$. The particle position and velocity are dynamic variables following the characteristic equations \eqref{eq:vlasov_char} and the weights are constant in time (note that time-dependent weights are also possible in advanced setups but are not discussed here). In order to compute the velocity moments of the distribution function needed to solve Maxwell's equations, a representation of the distribution function is necessary. This is usually given by
\begin{equation}
f_s(\xb, \vb, t) = \sum_{p=1}^{N_{p_s}} w_p S(\xb -\xb_p) \delta(\vb - \vb_p),
\end{equation}
where $S$ can either be a $\delta$ distribution or a smoothing kernel. Finally, the fields are represented on a grid and Maxwell's equations are solved by finite elements or finite differences.
Multiple discretization schemes have been discussed in the literature which have similar building blocks based on solutions of the field equations and loops over the particles with field evaluations for the particle push and current or charge depositions to assemble the source terms for the Maxwell's equations. Since usually the number of particles is much larger than the number of degrees of freedom in the description of the fields, the computational complexity is dominated by the particle loop.

The motivation of our work is an efficient and portable implementation of the scheme proposed in Ref.~\cite{GEMPIC} for the Vlasov--Maxwell equations. The scheme uses compatible spline finite elements as proposed by Buffa et al.~\cite{Buffa:2011} for the fields and a Klimontovic distribution for the particle distribution (i.e. $S = \delta$). Basis functions of different order are used to represent the magnetic and the electric field, respectively. Let us denote by $\Lambda^{1,k}_i$ the basis functions for component $k$, $k=1,2,3$, of the electric field associated with the grid point $i$, $i=1,\ldots, N_g$. In the same way, we denote by $\Lambda^{2,k}_i$ the basis functions for component $k$, $k=1,2,3$, of the magnetic field associated with grid point $i$, $i=1,\ldots, N_g$.
 Then, the semi-discretized fields are represented as
\begin{equation}
\tilde \Eb_k (\xb,t) = \sum_{i=1}^{N_g} e_{k,i}(t) \Lambda^{1,k}_i(\xb), \quad
\tilde \Bb_k (\xb,t) = \sum_{i=1}^{N_g} b_{k,i}(t) \Lambda^{2,k}_i(\xb).
\end{equation}
with $\eb_{k} = \left(e_{k,1}, \ldots, e_{k,N_g}\right)^\top$ and $\bb_k = \left(b_{k,1}, \ldots, b_{k,N_g}\right)^\top$ being the dynamic variables. Given a certain degree $p$, the basis functions are constructed in the following way:
\begin{itemize}
	\item $\Lambda^{1,k}$ is constructed as a tensor product of splines of degree $p-1$ in $x_k$ and of degree $p$ along the other two dimensions,
	\item $\Lambda^{2,k}$ is constructed as a tensor product of splines of degree $p$ in $x_k$ and of degree $p-1$ along the other two dimensions.
\end{itemize}
 Furthermore, equation \eqref{eq:maxwell_ampere} is solved in weak form and \eqref{eq:maxwell_faraday} is solved in strong form. Inserting the representation of the fields and particles into the equations yields the following semi-discrete equations of motion
\begin{subequations}\label{eq:vlasov_maxwell_equations_of_motion}
\begin{align}
\label{eq:vlasov_maxwell_equations_of_motion_x}
\frac{\du \X}{\du t}
&= \V ,
\\
\label{eq:vlasov_maxwell_equations_of_motion_v}
\frac{\du \V_1}{\du t}
&= \frac{q}{m} \left( \LaB^{1,1} (\X) \eb_1 + \V_2 \MM_q \LaB^{2,3} (\X) \bb_3 - \V_3 \MM_q \LaB^{2,2} (\X) \bb_2 \right) ,\\
\frac{\du \V_2}{\du t}
&= \frac{q}{m} \left( \LaB^{1,2} (\X) \eb_2 + \V_3 \MM_q \LaB^{2,1} (\X) \bb_1 - \V_1 \MM_q \LaB^{2,3} (\X) \bb_3 \right) ,\\
\frac{\du \V_3}{\du t}
&= \frac{q}{m} \left( \LaB^{1,3} (\X) \eb_3 + \V_1 \MM_q \LaB^{2,2} (\X) \bb_2 - \V_2 \MM_q \LaB^{2,1} (\X) \bb_1 \right) ,
\\
\label{eq:vlasov_maxwell_equations_of_motion_e}
\frac{\du \eb}{\du t}
&= \MM_{1}^{-1} \left( \C^\top \MM_{2} \bb (t) - \LaB^1 (\X)^\top \MM_q \V \right) ,
\\
\label{eq:vlasov_maxwell_equations_of_motion_b}
\frac{\du \bb}{\du t}
&= - \C \eb (t) .
\end{align}
\end{subequations}
The dynamic variables are given by $(\X, \V, \eb,\bb)$, where $\X = \left(\xb_1^\top, \ldots, \xb_{N_p}^\top \right)^\top$ and $\V = \left(\vb_1^\top, \ldots, \vb_{N_p}^\top \right)^\top$, $\eb = (\eb_1^\top, \eb_2^\top, \eb_3^\top)^\top$, $\bb = (\bb_1^\top, \bb_2^\top, \bb_3^\top)^\top$. Furthermore, $\C$ represents the discrete curl matrix, $\MM_j$ the finite element mass matrix for basis $j=1,2$, and $\MM_q$ is a diagonal matrix with the product of the particle weight and charge on the diagonals. The matrix $\LaB^j = (\LaB^{j,1}, \LaB^{j,2}, \LaB^{j,3})$ is a $N_p \times (3N_g)$ matrix with entries $(\LaB^{j,k})_{p,i} = \Lambda^{j,k}_i(\xb_p)$.

These equations can either be solved by some implicit time-stepping scheme or by a splitting of the equations into several subsystem that are chosen in such a way that the substeps can be solved explicitly. Such an explicit time stepping scheme, called Hamiltonian splitting, has been proposed in Ref.~\cite{GEMPIC}. For this study, we implement only one of the building blocks of the complete scheme that, however, contains the main features of the overall schemes. The building blocks solves the following part of the equations
\begin{subequations}
\begin{align}
\frac{\du \X_1}{\du t}
&= \V_1 ,\\
\frac{\du \V_2}{\du t}
&= -\frac{q}{m}   \V_1 \MM_q \LaB^{2,3} (\X) \bb_3  ,\\
\frac{\du \V_3}{\du t}
&= \frac{q}{m} \V_1 \MM_q \LaB^{2,2} (\X) \bb_2 ,\\
\MM_{1,1} \frac{\du \eb_1}{\du t} &= - \LaB^{1,1} (\X)^\top \MM_q \V_1,
\end{align}
\end{subequations}
by explicit integration over time as
\begin{subequations}
\begin{align}
%\X_1(t_{m+1}) = \X_1(t_m) + \Delta t \V_1(t_m),\label{eq:Hp1_int_x1}\\
%\V_2(t_{m+1}) = \V_2(t_m) - \frac{q}{m} \int_{t_m}^{t_{m+1}} \LaB_3^{2} \du \tau.
x_{1,p} (t_{m+1}) &= x_{1,p} (t_{m}) + \Delta t v_{1,p}, \label{eq:Hp1_int_x1}\\
v_{2,p} (t_{m+1}) &= v_{2,p} (t_{m}) - \frac{q}{m} \sum_{i=1}^{N_g} b_{3,i}(t_m) \int_{t_m}^{t_{m+1}} \Lambda^{2,3}(\xb_{p}(\tau)) v_{1,p}(t_m)\du \tau,\label{eq:Hp1_int_v2}\\
v_{3,p} (t_{m+1}) &= v_{3,p} (t_{m}) + \frac{q}{m} \sum_{i=1}^{N_g} b_{2,i}(t_m) \int_{t_m}^{t_{m+1}} \Lambda^{2,2}(\xb_{p}(\tau)) v_{1,p}(t_m)\du \tau,\label{eq:Hp1_int_v3}\\
\MM_{1,1}\eb_1(t_{m+1}) &=  \MM_{1,1}\eb_1(t_{m}) - q \sum_{p=1}^{N_p} w_p \int_{t_{m}}^{t_{m+1}} \LaB^{1,1}(\xb_p(\tau)) v_{1,p}(t_m)\du \tau.\label{eq:Hp1_int_e}
\end{align}
\end{subequations}
where $\xb_p( \tau) = \left( x_{1,p}(t_m) + \tau v_{1,p}(t_m), x_{2,p}(t_m), x_{3,p}(t_m) \right)^\top$.
The computationally most expensive part is to evaluate the integrals over the basis functions. If the velocity $v_{1,p}(t_m)$ is nonzero, we can transform the integral from $\tau$ to $\sigma = x_{1,p}(t_m) + \tau  v_{1,p}(t_m)$. Then equation \eqref{eq:Hp1_int_v2} reads
\begin{equation}
  v_{2,p} (t_{m+1}) = v_{2,p} (t_{m}) - \frac{q}{m} \sum_{i=1}^{N_g} b_{3,i}(t_m) \int_{x_{1,p}(t_m)}^{x_{1,p}(t_{m+1})} \Lambda^{2,3}(\sigma,  x_{2,p}(t_m), x_{3,p}(t_m))\du \sigma.
\end{equation}
The other integrals can be transformed in a similar way. Next, we apply the tensor product of one-variate splines $N^{q}(x)$ of degree $q=p$ and $q=p-1$ as proposed before. Then, the update of the particle velocities and the contribution of particle $p$ to component $i$ of the integrated current $j_i$ reads
\begin{subequations}\label{eq:Hp1_int_particle}
  \begin{align}
    v_{2,p} (t_{m+1}) &= v_{2,p} (t_{m}) - \frac{q}{m} \sum_{i=1}^{N_g} b_{3,i}(t_m) \int_{x_{1,p}(t_m)}^{x_{1,p}(t_{m+1})} N_{i_1}^{p-1}(\sigma)\du \sigma N_{i_1}^{p-1}( x_{2,p}(t_m)) N_{i_3}^{p}( x_{3,p}(t_m))\label{eq:Hp1_int_particle_v2}\\
    v_{3,p} (t_{m+1}) &= v_{3,p} (t_{m}) + \frac{q}{m} \sum_{i=1}^{N_g} b_{2,i}(t_m) \int_{x_{1,p}(t_m)}^{x_{1,p}(t_{m+1})} N_{i_1}^{p-1}(\sigma)\du \sigma N_{i_1}^{p}( x_{2,p}(t_m)) N_{i_3}^{p-1}( x_{3,p}(t_m)),\label{eq:Hp1_int_particle_v3}\\
    j_i &= j_i + q w_p \int_{x_{1,p}(t_m)}^{x_{1,p}(t_{m+1})} N_{i_1}^{p-1}(\sigma)\du \sigma N_{i_1}^{p}( x_{2,p}(t_m)) N_{i_3}^{p}( x_{3,p}(t_m))\label{eq:Hp1_int_particle_j1},
    \end{align}
\end{subequations}
where we decompose the unique index $i$ for the basis functions into a three dimensional index $(i_1,i_2,i_3)$ reflecting the tensor product structure of the basis. Since the basis functions have compact support, we identify first in which grid cell the particle is located and then we evaluate the $q+1$ basis functions (of degree $q$) with support on this cell. To evaluate the splines, we use their representation in piecewise polynomial form (pp-form) using Horner's algorithm at the particle position (normalized to the grid cell). For the evaluation of the integral, we evaluate the primitive basis function at the new and old position. This can either be implemented using numerical quadrature or based on the primitive function of $N^{p-1}$. We choose the latter approach and denoting the primitive of $N^{p-1}$ by $\mathcal{N}$, we get
\begin{equation}
  \int_{x_{1,p}(t_m)}^{x_{1,p}(t_{m+1})} N_{i_1}^{p-1}(\sigma)\du \sigma = \mathcal{N}_{i_1}(x_{1,p}(t_{m+1})) - \mathcal{N}_{i_1}(x_{1,p}(t_m)).
  \end{equation}
  We note that the support of a spline of degree $p-1$ is $p$ cells while the support of this integral is variable depending on how large the difference $x_{1,p}(t_{m+1})-x_{1,p}(t_{m})$ between the old and new particle position is. We note that this results in a variable loop length that makes it harder to design data structures for efficient and vectorized current deposition of this particular scheme. %Note that we concentrate on the particle loop solving \eqref{eq:Hp1_int_particle} for each particle and exclude the final step of updating the electric fields in \eqref{eq:Hp1_int_e} as this part only involves the degrees-of-freedom of the fields and is thus computationally much less demanding.

\section{Color-coded implementation comparison of the PIC routine}
\label{appendix:colorcoded}

The source code shown in listing \ref{code:color_hp1} contains the key
features necessary to implement the PIC routine for each parallel
programming model under investigation. In particular the color coding is as follows:
\textcolor{black}{Black for common code},
\textcolor{blue}{blue for OpenMP},
\textcolor{brown}{brown for OpenACC},
\textcolor{red}{red for Kokkos},
\textcolor{green}{green for RAJA},
\textcolor{purple}{purple for CUDA}, and
\textcolor{orange}{orange for hybrid OpenMP/Kokkos}.
Note that the listing presents a concatenation from the different source code
parts for illustration purposes.  %See the appendices \ref{appendix:kokkos} and
%\ref{appendix:raja} for the actual Kokkos and RAJA implementations.
%
\begin{lstlisting}[
caption={Color-coded key implementation features of the PIC routine for direct
comparison.},
label={code:color_hp1},
frame={tlrb}]
// Class particle group: Basic data structure saving
// the information on all particles of one species
|\textcolor{red}{class particle\_group\_base\_Kokkos \{}|
	|\textcolor{red}{KOKKOS\_FUNCTION particle\_group\_base\_Kokkos(char*, long, bool);}|
	// common particle weight
	double common_weight;
	|\textcolor{red}{ViewScalarDoubleType view\_common\_weight;}|
	|\textcolor{red}{ViewScalarDoubleType::HostMirror view\_common\_weight\_host;}|

	// array storing position in phase space and weight for each of n_particles
	double** particle_array;
	|\textcolor{red}{ViewVector2DType view\_particle\_array;}|
	|\textcolor{orange}{ViewVector2DUnmanagedType view\_particle\_array\_unmanaged;}|
	|\textcolor{red}{ViewVector2DType::HostMirror view\_particle\_array\_host;}|

	// information about the species ( mass and charge )
	double m; //mass of a single particle
	|\textcolor{red}{ViewScalarDoubleType view\_m;}|
	|\textcolor{orange}{ViewScalarDoubleUnmanagedType view\_m\_unmanaged;}|
	|\textcolor{red}{ViewScalarDoubleType::HostMirror view\_m\_host;}|

	//charge
	double q; //charge of a single particle
	|\textcolor{red}{ViewScalarDoubleType view\_q;}|
	|\textcolor{orange}{ViewScalarDoubleUnmanagedType view\_q\_unmanaged;}|
	|\textcolor{red}{ViewScalarDoubleType::HostMirror view\_q\_host;}|

	// number of particles in the particle group
	long n_particles;
	|\textcolor{red}{ViewScalarLongType view\_n\_particles;}|
	|\textcolor{orange}{ViewScalarLongUnmanagedType view\_n\_particles\_unmanaged;}|
	|\textcolor{red}{ViewScalarLongType::HostMirror view\_n\_particles\_host;}|
	|\textcolor{red}{// Next the accessor functions are defined}|
	|\textcolor{red}{\{ ... \}}|
|\textcolor{red}{\}}|

|\textcolor{green}{class particle\_group\_base\_RAJA \{}|
	|\textcolor{green}{particle\_group\_base\_RAJA(string, long, bool);}|
	// common particle weight
	double common_weight;

	// array storing position in phase space and weight for each of n_particles
	double** particle_array;
	|\textcolor{green}{double* d\_particle\_array;}|
	|\textcolor{green}{RajaVector2DType raja\_particle\_array;}|

	// information about the species ( mass and charge )
	double m; //mass of a single particle
	|\textcolor{green}{double *d\_m;}|
	|\textcolor{green}{RajaVector1DType raja\_m;}|

	double q; //charge of a single particle
	|\textcolor{green}{double *d\_q;}|
	|\textcolor{green}{RajaVector1DType raja\_q;}|

	// number of particles in the particle group
	long n_particles;
	|\textcolor{green}{long *d\_n\_particles;}|
	|\textcolor{green}{RajaVector1DLongType raja\_n\_particles;}|
	|\textcolor{green}{// Next the accessor functions are defined}|
	|\textcolor{green}{\{ ... \}}|
|\textcolor{green}{\}}|

|\textcolor{purple}{class particle\_group\_base\_cuda \{}|
	|\textcolor{purple}{particle\_group\_base\_cuda(string, long, bool);}|
	// common particle weight
	double common_weight;

	// array storing position in phase space and weight for each of n_particles
	double** particle_array;
	|\textcolor{purple}{double* d\_particle\_array;}|

	// information about the species ( mass and charge )
	double m; //mass of a single particle
	|\textcolor{purple}{double *d\_m;}|

	double q; //charge of a single particle
	|\textcolor{purple}{double *d\_q;}|

	// number of particles in the particle group
	long n_particles;
	|\textcolor{purple}{long *d\_n\_particles;}|
	|\textcolor{purple}{// Next the accessor functions are defined}|
	|\textcolor{purple}{\{ ... \}}|
|\textcolor{purple}{\}}|

void hamiltonian_splitting::pic_routine(double dt,
					|\textcolor{red}{int n\_threads, int n\_teams}|
					|\textcolor{green}{int n\_threads}|
					|\textcolor{purple}{int n\_threads}|) {
	|\textcolor{blue}{int total\_num\_threads = omp\_get\_max\_threads();}|
	|\textcolor{blue}{//OpenMP: initialisation of the reduction array}|
	|\textcolor{blue}{double j\_dofs[total\_num\_threads][this->part\_mesh\_coupling.n\_dofs];}|
	|\textcolor{blue}{for(int i=0; i<total\_num\_threads; i++) \{}|
		|\textcolor{blue}{for(int j=0; j<this->part\_mesh\_coupling.n\_dofs; j++) \{}|
			|\textcolor{blue}{j\_dofs[i][j] = 0.0;}|
		|\textcolor{blue}{\}}|
	|\textcolor{blue}{\}}|

	|\textcolor{red}{//Kokkos: initialisation of the reduction array}|
	|\textcolor{red}{for(long i=0; i<this->view\_j\_dofs\_local\_host.size(); i++) \{}|
		|\textcolor{red}{this->view\_j\_dofs\_local\_host(i) = 0.0;}|
	|\textcolor{red}{\}}|
	|\textcolor{red}{Kokkos::deep\_copy(this->view\_j\_dofs\_local, this->view\_j\_dof\_local\_host);}|

	|\textcolor{green}{int total\_num\_threads = std::max(atoi(std::getenv("OMP\_NUM\_THREADS")), 1);}|
	|\textcolor{green}{//RAJA: initialisation of the reduction array}|
	|\textcolor{green}{\#if defined(RAJA\_ENABLE\_CUDA) \&\& defined(I\_USE\_CUDA)}|
		|\textcolor{green}{RAJA::forall<RAJA::cuda\_exec<256{>}>(RAJA::RangeSegment(0, this->part\_mesh\_coupling.n\_dofs),}|
		 |\textcolor{green}{[*this] RAJA\_DEVICE (int i) \{}|
			|\textcolor{green}{if(i<this->part\_mesh\_coupling.raja\_n\_dofs(0)) \{}|
				|\textcolor{green}{this->raja\_j\_dofs\_local(i) = 0.0;}|
			|\textcolor{green}{\}}|
		 |\textcolor{green}{\});}|
	|\textcolor{green}{\#else}|
		|\textcolor{green}{RAJA::forall<RAJA::omp\_parallel\_for\_exec>(RAJA::RangeSegment(0, this->part\_mesh\_coupling.n\_dofs),}|
		 |\textcolor{green}{[this] (int i) \{}|
			|\textcolor{green}{if(i<part\_mesh\_coupling.raja\_n\_dofs(0)) \{}|
				|\textcolor{green}{raja\_j\_dofs\_local(i) = 0.0;}|
			|\textcolor{green}{\}}|
		 |\textcolor{green}{\});}|
	|\textcolor{green}{\#endif}|

	|\textcolor{purple}{int M\_blocks = (this->particle\_group.group[0].n\_particles + n\_threads-1)/n\_threads;	//\#of blocks}|
	|\textcolor{purple}{//CUDA: no reduction array, atomic operations are used instead}|

	|\textcolor{orange}{int total\_num\_threads = omp\_get\_max\_threads();}|
	|\textcolor{orange}{//Hybrid: initialisation of the reduction array}|
	|\textcolor{orange}{double j\_dofs[total\_num\_threads][this->part\_mesh\_coupling.view\_n\_dofs\_host(0)];}|
	|\textcolor{orange}{for(int i=0; i<total\_num\_threads; i++) \{}|
		|\textcolor{orange}{for(int j=0; j<this->part\_mesh\_coupling.view\_n\_dofs\_host(0); j++) \{}|
			|\textcolor{orange}{j\_dofs[i][j] = 0.0;}|
		|\textcolor{orange}{\}}|
	|\textcolor{orange}{\}}|
	|\textcolor{orange}{//Hybrid: set all unmanaged views}|
	|\textcolor{orange}{this->view\_i\_weight\_unmanaged = this->view\_i\_weight;}|
	|\textcolor{orange}{\{...\}}|

	|\textcolor{brown}{int total\_num\_gangs = input\_n\_gangs();}|
	|\textcolor{brown}{int total\_num\_vectors = input\_n\_vectors();}|
	|\textcolor{brown}{//OpenACC: initialisation of the reduction array}|
	|\textcolor{brown}{double **j\_dofs = (double**)malloc(total\_num\_gangs * sizeof(double *));}|
	|\textcolor{brown}{for(int i=0; i<total\_num\_gangs; i++) \{}|
		|\textcolor{brown}{j\_dofs[i] = (double*)malloc(this->part\_mesh\_coupling.n\_dofs * sizeof(double));}|
		|\textcolor{brown}{for(int j=0; j<this->part\_mesh\_coupling.n\_dofs; j++) \{}|
			|\textcolor{brown}{j\_dofs[i][j] = 0.0;}|
		|\textcolor{brown}{\}}|
	|\textcolor{brown}{\}}|

	|\textcolor{blue}{\#pragma omp parallel}|
	|\textcolor{blue}{\{}|
		|\textcolor{blue}{//OpenMP: initialisation of the local variables}|
		|\textcolor{blue}{\{...\}}|
		|\textcolor{blue}{for(long i\_sp = 0; i\_sp<this->particle\_group.n\_species; i\_sp++) \{}|
			|\textcolor{blue}{//OpenMP: get thread ID, set qoverm}|
			|\textcolor{blue}{\{...\}}|
			|\textcolor{blue}{\#pragma omp for}|
			|\textcolor{blue}{for(long i\_part=0; i\_part<this->particle\_group.group[i\_sp].n\_particles; i\_part++) \{}|
				|\textcolor{blue}{\{operator\_openmp\}}|
			|\textcolor{blue}{\}}|
			|\textcolor{blue}{\#pragma omp for}|
			|\textcolor{blue}{for(int i=0; i<this->part\_mesh\_coupling.n\_dofs; i++) \{}|
				|\textcolor{blue}{for(int thread=1; thread<total\_num\_threads; thread++) \{}|
					|\textcolor{blue}{j\_dofs[0][i] += j\_dofs[thread][i];}|
				|\textcolor{blue}{\}}|
			|\textcolor{blue}{\}}|
		|\textcolor{blue}{\}}|
	|\textcolor{blue}{\}}|

	|\textcolor{red}{//Kokkos: set dt}|
	|\textcolor{red}{\{...\}}|
	|\textcolor{red}{for(long i\_sp = 0; i\_sp<this->particle\_group.view\_n\_species\_host(0); i\_sp++) \{}|
		|\textcolor{red}{//Kokkos: set view\_i\_sp and shared\_size=n\_shared*sizeof(double)}|
		|\textcolor{red}{\{...\}}|
		|\textcolor{red}{//Kokkos: set parallel policy}|
		|\textcolor{red}{auto policy = Kokkos::TeamPolicy<pic\_routine, ExecSpace>((this->particle\_group.view\_group\_host(i\_sp).n\_particles + n\_threads-1)/n\_threads, n\_threads)}|
						|\textcolor{red}{.set\_scratch\_size(1,Kokkos::PerThread(shared\_size));}|
		|\textcolor{red}{//Kokkos: parallel computation}|
		|\textcolor{red}{Kokkos::parallel\_for(policy, *this);}|
	|\textcolor{red}{\}}|

	|\textcolor{green}{//RAJA: set dt}|
	|\textcolor{green}{\{...\}}|
	|\textcolor{green}{for(long i\_sp = 0; i\_sp<this->particle\_group.n\_species; i\_sp++) \{}|
		|\textcolor{green}{//RAJA: set view\_i\_sp and shared\_size}|

		|\textcolor{green}{\#if defined(RAJA\_ENABLE\_CUDA) \&\& defined(I\_USE\_CUDA)}|
		|\textcolor{green}{//RAJA: GPU parallel computation}|
		|\textcolor{green}{RAJA::forall<RAJA::cuda\_exec<256{>}>(RAJA::RangeSegment(0, this->particle\_group.group[0].n\_particles),}|
			|\textcolor{green}{[=,*this] RAJA\_DEVICE (int i\_part) \{}|
				|\textcolor{green}{\{operator\_RAJA\_GPU\}	//RAJA GPU: uses atomics instead of reduction}|
			|\textcolor{green}{\});}|
		|\textcolor{green}{\#else}|
		|\textcolor{green}{//RAJA CPU: raja\_private\_raja\_array is our solution to false-sharing}|
		|\textcolor{green}{//It is an array of struct such that each thread gets contiguous memory to use as scratch}|

		|\textcolor{green}{//RAJA: set parallel policy}|
		|\textcolor{green}{using fdPolicy = RAJA::KernelPolicy< RAJA::statement::For< 0, RAJA::omp\_parallel\_for\_exec, RAJA::statement::Lambda<0> > >;}|

		|\textcolor{green}{//RAJA CPU: initialisation of the handmade reduction 2D array}|
		|\textcolor{green}{RAJA::kernel<fdPolicy>(RAJA::make\_tuple(RAJA::RangeSegment(0, total\_num\_threads)),}|
			|\textcolor{green}{[\&,this] (int i\_part) \{}|
				|\textcolor{green}{for(int i=0; i<this->part\_mesh\_coupling.n\_dofs; i++)}|
				 |\textcolor{green}{this->raja\_private\_raja\_array(i\_part).raja\_handmade\_reduce(i) = 0.0;}|
			|\textcolor{green}{ \});}|


		|\textcolor{green}{//RAJA: CPU parallel computation}|
		|\textcolor{green}{RAJA::kernel<fdPolicy>(RAJA::make\_tuple(RAJA::RangeSegment(0, this->particle\_group.group[0].n\_particles)),}|
			|\textcolor{green}{[\&,this] (int i\_part) \{}|
				|\textcolor{green}{\{operator\_RAJA\_CPU\}}|
			 |\textcolor{green}{\});}|

		|\textcolor{green}{//RAJA: CPU handmade reduction}|
		|\textcolor{green}{RAJA::kernel<fdPolicy>(RAJA::make\_tuple(RAJA::RangeSegment(0, this->part\_mesh\_coupling.n\_dofs)),}|
			 |\textcolor{green}{[\&,this] (int i\_part) \{}|
			 |\textcolor{green}{for(int i=0; i<total\_num\_threads; i++)}|
				|\textcolor{green}{this->d\_j\_dofs\_local[i\_part] += this->raja\_private\_raja\_array(i).raja\_handmade\_reduce(i\_part);}|
		|\textcolor{green}{\});}|
		|\textcolor{green}{\#endif}|
	|\textcolor{green}{\}}|

	|\textcolor{purple}{//CUDA: parallel computation}|
	|\textcolor{purple}{operator\_cuda<{<}<M\_blocks,n\_threads>{>}>(0.05, 0, this->d\_particle\_group, this->d\_part\_mesh\_coupling,}|
							 |\textcolor{purple}{this->d\_control\_vari, this->i\_weight, this->d\_bfield\_dofs, this->d\_j\_dofs\_local, this->d\_Lx);}|

	|\textcolor{orange}{\#pragma omp parallel}|
	|\textcolor{orange}{\{}|
		|\textcolor{orange}{//Hybrid: initialisation of the local variables}|
		|\textcolor{orange}{\{...\}}|
		|\textcolor{orange}{for(long i\_sp = 0; i\_sp<this->particle\_group.view\_n\_species\_unmanaged(0); i\_sp++) \{}|
			|\textcolor{orange}{//Hybrid: get thread ID, set qoverm}|
			|\textcolor{orange}{\{...\}}|
			|\textcolor{orange}{\#pragma omp for}|
			|\textcolor{orange}{for(long i\_part=0; i\_part<this->particle\_group.view\_group\_unmanaged( this->view\_i\_species\_unmanaged(0) ).view\_n\_particles\_unmanaged(0); i\_part++) \{}|
				|\textcolor{orange}{\{operator\_hybrid\}}|
			|\textcolor{orange}{\}}|
			|\textcolor{orange}{\#pragma omp for}|
			|\textcolor{orange}{for(int i=0; i<this->part\_mesh\_coupling.view\_n\_dofs\_unmanaged(0); i++) \{}|
				|\textcolor{orange}{for(int thread=1; thread<total\_num\_threads; thread++) \{}|
					|\textcolor{orange}{j\_dofs[0][i] += j\_dofs[thread][i];}|
				|\textcolor{orange}{\}}|
			|\textcolor{orange}{\}}|
		|\textcolor{orange}{\}}|
	|\textcolor{orange}{\}}|

	|\textcolor{brown}{\#pragma acc data}|
		|\textcolor{brown}{copy(j\_dofs[0:total\_num\_gangs][0:this->part\_mesh\_coupling.n\_dofs])}|
		|\textcolor{brown}{copyin(this[0:1],}|
			|\textcolor{brown}{this->part\_mesh\_coupling,}|
			|\textcolor{brown}{this->part\_mesh\_coupling.domain[0:3][0:2],}|
			|\textcolor{brown}{this->part\_mesh\_coupling.delta\_x[0:3],}|
			|\textcolor{brown}{this->part\_mesh\_coupling.rdelta\_x[0:3],}|
			|\textcolor{brown}{this->part\_mesh\_coupling.spline\_0,}|
			|\textcolor{brown}{this->part\_mesh\_coupling.spline\_0.d\_poly\_coeffs[0:(spline\_0\_degree+1)*(spline\_0\_degree+1)],}|
			|\textcolor{brown}{this->part\_mesh\_coupling.spline\_0.d\_poly\_coeffs\_fpa[0:(spline\_0\_degree+2)*(spline\_0\_degree+1)],}|
			|\textcolor{brown}{this->part\_mesh\_coupling.spline\_1,}|
			|\textcolor{brown}{this->part\_mesh\_coupling.spline\_1.d\_poly\_coeffs[0:(spline\_1\_degree+1)*(spline\_1\_degree+1)],}|
			|\textcolor{brown}{this->part\_mesh\_coupling.spline\_1.d\_poly\_coeffs\_fpa[0:(spline\_1\_degree+2)*(spline\_1\_degree+1)],}|
			|\textcolor{brown}{this->part\_mesh\_coupling.n\_grid[0:3],}|
			|\textcolor{brown}{this->part\_mesh\_coupling,}|
			|\textcolor{brown}{this->Lx[0:3],}|
			|\textcolor{brown}{this->bfield\_dofs[0:this->part\_mesh\_coupling.n\_dofs*3],}|
			|\textcolor{brown}{this->particle\_group,}|
			|\textcolor{brown}{this->particle\_group.group[0:this->particle\_group.n\_species],}|
			|\textcolor{brown}{this->particle\_group.group[0].sp),}|
		|\textcolor{brown}{copy(  this->particle\_group.group[0].particle\_array[0:7][0:this->particle\_group.group[0].n\_particles])}|
	|\textcolor{brown}{\{}|
		|\textcolor{brown}{\#pragma acc parallel num\_gangs(total\_num\_gangs) vector\_length(total\_num\_vectors)}|
		|\textcolor{brown}{\{}|
			|\textcolor{brown}{//OpenACC: initialisation of the per-gang reduction j\_dofs, local variables and get gang ID}|
			|\textcolor{brown}{int this\_gang = \_\_pgi\_gangidx();}|
			|\textcolor{brown}{\{...\}}|
			|\textcolor{brown}{for(long i\_sp = 0; i\_sp<this->particle\_group.n\_species; i\_sp++) \{}|
				|\textcolor{brown}{//OpenACC: set qoverm}|
				|\textcolor{brown}{\{...\}}|
				|\textcolor{brown}{\#pragma acc loop}|
				|\textcolor{brown}{for(long i\_part=0; i\_part<this->particle\_group.group[i\_sp].n\_particles; i\_part++) \{}|
					|\textcolor{brown}{\{operator\_openacc\}}|
				|\textcolor{brown}{\}}|
			|\textcolor{brown}{\}}|
		|\textcolor{brown}{\}}|
	|\textcolor{brown}{\}}|

	|\textcolor{blue}{//OpenMP: copying the reduction's result}|
	|\textcolor{blue}{for(int i=0; i<this->part\_mesh\_coupling.n\_dofs; i++) \{}|
		|\textcolor{blue}{this->j\_dofs\_local[i] = j\_dofs[0][i];}|
	|\textcolor{blue}{\}}|

	|\textcolor{orange}{//Hybrid: copying the reduction's result}|
	|\textcolor{orange}{for(int i=0; i<this->part\_mesh\_coupling.view\_n\_dofs(0); i++) \{}|
		|\textcolor{orange}{this->view\_j\_dofs\_local(i) = j\_dofs[0][i];}|
	|\textcolor{orange}{\}}|

	|\textcolor{brown}{//OpenACC: second reduction on the gang results}|
	|\textcolor{brown}{for(int i=0; i<total\_num\_gangs; i++) \{}|
		|\textcolor{brown}{for(int j=0; j<this->part\_mesh\_coupling.n\_dofs; j++) \{}|
			|\textcolor{brown}{this->j\_dofs\_local[j] += j\_dofs[i][j];}|
		|\textcolor{brown}{\}}|
	|\textcolor{brown}{\}}|
	
	|\textcolor{brown}{//OpenACC: free memory}|
	|\textcolor{brown}{for(int i=0; i<total\_num\_gangs; i++)}|
		|\textcolor{brown}{free(j\_dofs[i]);}|
	|\textcolor{brown}{free(j\_dofs);}|
}

operator_openmp {
|\textcolor{red}{KOKKOS\_FUNCTION void hamiltonian\_splitting::operator() (const pic\_routine\&, const Kokkos::TeamPolicy<>::member\_type \& team\_member) const \{}|
|\textcolor{green}{operator\_RAJA\_GPU \{}|
|\textcolor{green}{operator\_RAJA\_CPU \{}|
|\textcolor{purple}{\_\_global\_\_ void operator\_cuda(double dt, int i\_sp, particles *d\_particle\_group, particle\_mesh\_coupling *d\_part\_mesh\_coupling, control\_variate *d\_control\_vari, long i\_weight, double *d\_bfield\_dofs, double *d\_j\_dofs, double *d\_Lx) \{}|
|\textcolor{orange}{operator\_hybrid \{}|
|\textcolor{brown}{operator\_openacc \{}|
	|\textcolor{red}{//Kokkos: thread access to special scatter view, used for vector reduction}|
	|\textcolor{red}{ViewScatterAccessType scatter\_access = this->scatter\_view.access();}|

	|\textcolor{red}{//Kokkos: initialise scratch variables}|
	|\textcolor{red}{ViewVector3ScratchType view\_x\_old(team\_member.thread\_scratch(1));}|
	|\textcolor{red}{\{...\}}|

	|\textcolor{green}{//RAJA GPU: initialisation of the local variables	BUT WITH FIXED SIZE}|
	|\textcolor{green}{//The size is given by the number of dimensions (3) by the support of the spline (4 here for cubic splines, but should be templated for varying order).}|
	|\textcolor{green}{double d\_spline\_0[4*3];}|
	|\textcolor{green}{\{...\}}|

	|\textcolor{purple}{//CUDA: initialisation of the local variables (and qoverm)	BUT WITH FIXED SIZE}|
	|\textcolor{purple}{double d\_spline\_0[4*3]; // Size as for RAJA.}|
	|\textcolor{purple}{\{...\}}|
	
	|\textcolor{brown}{//OpenACC: initialisation of the local variables (and qoverm)	BUT WITH FIXED SIZE}|
	|\textcolor{brown}{double d\_spline\_0[4*3]; // Size as for RAJA.}|
	|\textcolor{brown}{\{...\}}|

	//Read out particle position and velocity
	{...}
	//Then update particle position:	X_new(0) = X_old(0) + dt * V(0)
	x_new[0] = x_old[0] + dt * vi[0];
	x_new[1] = x_old[1];
	x_new[2] = x_old[2];
	//Get charge for accumulation of j
	{...}

	|\textcolor{blue}{this->part\_mesh\_coupling.add\_current\_update\_v\_primitive\_component1\_spline\_3d\_feec\_util (res[this\_thread], x\_old, x\_new[0], wi[0], qoverm,}|
			|\textcolor{blue}{this->bfield\_dofs, vi, \&util\_arrays);}|

	|\textcolor{red}{this->part\_mesh\_coupling.view\_add\_current\_update\_v\_primitive\_component1\_spline\_3d\_feec\_scratch (team\_member,}|
						|\textcolor{red}{view\_x\_old, view\_x\_new(0), view\_wi(0), qoverm,}|
						|\textcolor{red}{this->view\_bfield\_dofs, view\_vi, \&(this->scatter\_view));}|

	|\textcolor{green}{this->part\_mesh\_coupling.raja\_add\_current\_update\_v\_primitive\_component1\_spline\_3d\_feec\_util (raja\_x\_old, raja\_x\_new(0), raja\_wi(0), qoverm,}|
	|\textcolor{green}{this->raja\_bfield\_dofs, raja\_vi, this->d\_j\_dofs\_local, \&util\_raja);}|	//GPU

	|\textcolor{green}{this->part\_mesh\_coupling.raja\_add\_current\_update\_v\_primitive\_component1\_spline\_3d\_feec\_util\_pool\_thread<0>}|
	|\textcolor{green}{(THREAD, this->raja\_private\_raja\_array(THREAD).raja\_x\_old,}|
	|\textcolor{green}{this->raja\_private\_raja\_array(THREAD).raja\_x\_new(0), this->raja\_private\_raja\_array(THREAD).raja\_wi(0),}|
	|\textcolor{green}{qoverm, this->raja\_bfield\_dofs, this->raja\_private\_raja\_array(THREAD).raja\_vi,}|
	|\textcolor{green}{this->raja\_private\_raja\_array(THREAD).raja\_handmade\_reduce, this->raja\_private\_raja\_array(THREAD));}|	//CPU

	|\textcolor{purple}{d\_part\_mesh\_coupling->cuda\_add\_current\_update\_v\_primitive\_component1\_spline\_3d\_feec\_util (x\_old, x\_new[0], wi[0], qoverm,}|
	|\textcolor{purple}{d\_bfield\_dofs, vi, d\_j\_dofs, \&util\_cuda);}|

	|\textcolor{orange}{this->part\_mesh\_coupling.add\_current\_update\_v\_primitive\_component1\_spline\_3d\_feec\_util\_from\_view (j\_dofs[this\_thread], x\_old, x\_new[0], wi[0], qoverm,}|
			|\textcolor{orange}{\&this->view\_bfield\_dofs\_unmanaged, vi, \&util\_arrays);}|

	|\textcolor{brown}{this->part\_mesh\_coupling.add\_current\_update\_v\_primitive\_component1\_spline\_3d\_feec\_util\_openacc (j\_dofs[this\_gang], x\_old, x\_new[0], wi[0], qoverm,}|
			|\textcolor{brown}{this->bfield\_dofs, vi, \&util\_openacc);}|
						
	x_new[0] = fmod(x_new[0] + this->Lx[0], this->Lx[0]);
	this->particle_group.group[i_sp].set_x(i_part, x_new);
	this->particle_group.group[i_sp].set_v(i_part, vi);
}

void particle_mesh_coupling::add_current_update_v_primitive_component1_spline_3d_feec_util (double *j_dofs,
			double *position_old, double position_new, double marker_charge, double qoverm,
			double *bfield_dofs,
			|\textcolor{orange}{ViewVector1DUnmanagedType *view\_bfield\_dofs\_unmanaged,}|	//instead of double *bfield_dofs
			double *vi, struct private_arrays *util_arrays,
			|\textcolor{green}{struct private\_raja util\_raja}|	//GPU
			|\textcolor{green}{struct private\_raja * util\_raja}|	//CPU
			|\textcolor{purple}{struct private\_cuda util\_cuda}|
			|\textcolor{brown}{struct private\_openacc util\_openacc}|
			) {
	//Initialise local variables (and qoverm)	BUT WITH FIXED SIZE
	{...}

	|\textcolor{red}{ViewScatterAccessType view\_j\_dofs\_scatter\_access = view\_j\_dofs\_scatter->access();}|
	// Identify the grid cell (box) where the particle is located and
	// its normalized position (xi) within the box
        {...}
	// Similarly as above, we identify box and normalized position for the
	// new position (along x(component) only)
        {...}
	// Extract box index along the direction component for the old position
        {...}
	// In the tensor product basis, the three 1D components are
	// combined with each other in every possible combination
	// Therefore, we start by computing the three 1D components
        {...}
	// Along x(0), we have to integrate over splines of degree degree-1
	// We get a contribution to the current in all indices of
	// splines that are nonzero in either boxnew or boxold
	// In each box, degree basis function are nonzero
	// This gives the following number of nonzero values of
	// the total integral
        {...}
	// First we evaluate the primitive of the degree basis functions that are
	// nonzero at the old and new particle positions respectively
	// For this we use the pp coefficients of the primitive function
	// of the splines and evaluate using Horner's scheme
        {...}
	// Now, we glue everything together
	// Note that the primitive function is equal to delta_x(component)
	// in all intervals with index larger that the indices of the
	// support of the basis function
        {...}
	// For the other two other directions, we need to evaluate the splines of
	// degree p and (p-1) at the particle position
        {...}
	// We use the pp form and evaluate with Horner's scheme
        {...}
	// Set index of first basis function that is nonzero at
	// the particle position

	// Set index of first basis function that is nonzero at
	// the particle position for the dimension with integration
        {...}
	//optimisation : precomputations of the indices
	{...}

	// loop over all the basis functions that are nonzero at the
	// particle position and update velocities and current
	for(long k=0; k<this->spline_degree+1; k++) {
		vtt2 = 0.0; vtt3 = 0.0;
		for(long j=0; j<this->spline_degree+1; j++) {
			// Save the 2D spline product to avoid recomputation in the inner loop
			splinejk = util_arrays->spline_0[1][j] * util_arrays->spline_0[2][k] * marker_charge;
			util_arrays->vt[0] = 0.0; util_arrays->vt[1] = 0.0;

			for(long i=0; i<local_size; i++) {
				// Compute 1D index of the basis function from the 3D tensor product index
				index1d = util_arrays->startjk[k][j] + util_arrays->index_x[i];
				// update the current
				|\textcolor{blue}{j\_dofs[index1d] += util\_arrays->j1d[i] * splinejk;}|
				|\textcolor{red}{view\_j\_dofs\_scatter\_access(index1d) += view\_j1d(i) * splinejk;}|
				|\textcolor{green}{RAJA::atomic::atomicAdd<RAJA::atomic::cuda\_atomic>(\&(d\_j\_dofs\_tmp[index1d]), util\_raja->raja\_j1d(i) * splinejk);}|	//GPU
				|\textcolor{green}{d\_j\_dofs\_tmp(index1d) += util\_raja.raja\_j1d(i) * splinejk;}|	//CPU
				|\textcolor{purple}{atomicAdd(\&(d\_j\_dofs\_tmp[index1d]), util\_cuda->d\_j1d(i) * splinejk);}|
				|\textcolor{orange}{j\_dofs[index1d] += util\_arrays->j1d[i] * splinejk;}|
				|\textcolor{brown}{\#pragma acc atomic}|
				|\textcolor{brown}{j\_dofs[index1d] += util\_openacc->d\_j1d[i] * splinejk;}|
				// contributions for the velocities
				util_arrays->vt[0] += bfield_dofs[start1+index1d] * util_arrays->j1d[i];
				util_arrays->vt[1] += bfield_dofs[start2+index1d] * util_arrays->j1d[i];
			}

			if(j>0) {
				vtt2 += util_arrays->vt[0]*util_arrays->spline_1[1][j-1];
			}
			vtt3 -= util_arrays->vt[1]*util_arrays->spline_0[1][j];
		}
		// update the velocities
		vi[1] -= qoverm*vtt2*util_arrays->spline_0[2][k];
		if(k>0) {
			vi[2] -= qoverm*vtt3*util_arrays->spline_1[2][k-1];
		}
	}
}
\end{lstlisting}

\end{document}